\documentstyle[art11,epsf]{article}
\def\sez#1{\section{#1} \setcounter{equation}{0}}
\def\cap#1{\caption[#1]{\it #1}}

\def\ev{{\it e\hskip-0.18em V}}
\def\kev{{\it ke\hskip-0.18em V}}
\def\mev{{\it Me\hskip-0.18em V}}

\def\lapprox{\mathrel{\mathop
  {\hbox{\lower0.5ex\hbox{$\sim$}\kern-0.8em\lower-0.7ex\hbox{$<$}}}}}
\def\gapprox{\mathrel{\mathop
  {\hbox{\lower0.5ex\hbox{$\sim$}\kern-0.8em\lower-0.7ex\hbox{$>$}}}}}
\def\papprox{\mathrel{\mathop
  {\hbox{\lower0.5ex\hbox{$\propto$}\kern-0.8em\lower-0.7ex\hbox{$\sim$}}}}}
\def\mb#1{\mbox{$#1$}}
\def\abs#1{\left| #1 \right|}
\def\nue{\nu_e}
\def\anue{\bar\nu_e}
\def\num{\nu_{\mu}}
\def\nut{\nu_{\tau}}
\def\nui{\nu_i}

\def\nux{\nu_x}

\def\mnut{m}
\def\anut{\alpha_{\nu_i}}
\def\eps{\epsilon _\nu}
\def\ee{\epsilon _e}
\def\emin{\epsilon _{{\it min}}}
\def\emax{\epsilon _{{\it max}}}
\def\thf{\vartheta _{{\it fw}}}
\def\tt{t_{{\it tr}}}

\begin{document}

\title{Supernova neutrinos and the $\nut$ mass.}
\author{Gianni {\sc Fiorentini}$^{1,2}$ and Camillo {\sc Acerbi}$^1$ 
\\[1.1em] 
\small 
$^1$ Istituto Nazionale di Fisica Nucleare - Sezione di Ferrara \\
\small 
$^2$ Dipartimento di Fisica - Universit\`a di Ferrara$\hskip 45pt$ \\[0.5em] 
\small 12, via Paradiso -- I-44100 Ferrara -- {\sc Italy} \\[1.0em]}
\date{December 23, 1996}
\maketitle

\begin{abstract}

We perform an extensive investigation of the sensitivity to non-vanishing 
$\nut$ mass in a large water \v{C}erenkov detector, developing an analysis 
method for neutrino events originated by a supernova explosion. This approach, 
based on directional considerations, provides informations almost undepending 
on the supernova model.
We analyze several theoretical models from numerical simulations and 
phenomenological models based on $SN1987A$ data, and determine optimal values 
of the analysis parameters so as to reach the highest sensitivity to a 
non-vanishing $\nut$ mass. The minimal detectable mass is generally just above 
the cosmologically interesting range, $m\sim 100\:\ev$, in the case of a 
supernova explosion near the galactic center. For the case that no positive 
signal is obtained, observation of a neutrino burst with Super-Kamiokande will 
anyhow lower the present upper bound on $\nut$ mass to few hundred $\ev$.

\end{abstract}

\pagestyle{plain}

\sez{Introduction.}

Upper bounds on the electron (anti-) neutrino mass at the level of few
$\ev$ are obtained from $\beta$ decay experiments \cite{pdg}, and have
been confirmed by data on neutrinos from $SN1987A$, collected by Kamiokande-II 
\cite{kII} and IMB \cite{imb} detectors, see e.g. \cite{lim_sn,l&l}.
Concerning the masses of $\num$ and $\nut$, the present experimental bounds 
are much less compelling: \mb{m_{\num}\lapprox 170\: \kev},
\mb{\mnut\lapprox 24\: \mev} \cite{pdg}. Therefore, any approach 
capable of significantly lowering these limits is highly 
desirable. Particularly, an observation sensitive to neutrino masses in 
the range \mb{10 - 100\: \ev} would be extremely important for cosmological
implications, see \cite{gershtein}.
Previous analyses \cite{seck,cline} claimed that this goal might be reached
with new generation larger detectors, for a supernova near the galactic 
center.

The questions we address in this paper are the following:
\begin{enumerate}
\item Which assumptions about neutrino emission are critical for
observing the effects of a non-vanishing mass?
\item How the analysis procedure can be optimized in order to reach the highest 
sensitivity to neutrino masses?
\item Which masses for $\num$ and $\nut$ can be actually explored with a 
detector such as Super-Kamiokande?
\end{enumerate}

In general, the determination of $\nu$-masses by means of a supernova 
observation is based on the comparison between the detected signal 
and the expected one for the case $m=0$.
A non-vanishing mass, in fact, would reflect in an overall delay of the events 
number distribution, and also in a spectral distortion of the signal, due to 
the energy dependence of the flight-time for massive neutrinos:
\equation \Delta t\, (\eps) =  {D\over  2c}\left( {m c^2 \over \eps}
\right) ^2 \endequation
where $D$ is the supernova distance, $c$ is the light speed, and $\eps$ is
the neutrino energy. 
The greatest difficult in applying this procedure is in the dependence 
on the supernova model assumed to calculate the expected signal. 
Moreover, when interested in $\num$ and $\nut$ (hereafter we will refer to 
these neutrinos, and their antiparticles, as $\nui$), there is an additional 
problem: it is impossible to isolate their signal from that of $\nue$ and
$\anue$, because all $\nui$ detection processes (namely neutral current 
reactions) are allowed for electron neutrinos too.

On the other hand, charged current interactions (only possible for $\nue$ and 
$\anue$ at the energies of interest) might be used to clearly identify part of 
the electron neutrino signal. 
One can thus conceive the following strategy:
\begin{enumerate}
\item split, by use of a distinctive signature, the overall set 
of events in two classes: the first one containing only $\nue$ and $\anue$, 
the second one involving all  neutrino species;
\item use the first class of events to derive the expected distribution 
for $\nue$ and $\anue$. 
In view of the strict limit on $m_{\nue}$, we can treat these neutrinos 
as massless for our purpose (e.g. from \mb{m_{\nue}\lapprox 15\: \ev} 
\cite{pdg} and equation 1.1, we get \mb{\Delta t \sim {\cal O}(1\, s)} 
even for the lowest detectable energies, \mb{\eps\simeq 5\:\mev}, for a SN near 
the galactic center);
\item use this distribution to infer that of $\num$ and $\nut$, again for
massless neutrinos;
\item compare the observed signal with that theoretically built in the previous 
step and look for mass effects. We assume that only one neutrino (say $\nut$) 
is massive and that flavor oscillations do not occur (see \cite{msw} and 
references therein for search of $\nut$ mass by use of MSW effects). 
The comparison should be performed in the second class of events 
(\mb{\nue +\nui}): in such a way we deal with a subset which still 
includes the whole $\nui$ signal, but only a part of the $\anue$ ``background''.
\end{enumerate}

The residual dependence on the supernova model is essentially in step 3: how
to connect the $\nue$ and $\anue$ emission features with those of 
$\nui$. In this paper we extensively investigate this point, by 
analyzing several theoretical and phenomenological SN models and their 
implications for the extraction of a non-vanishing $\nut$ mass.

We also discuss how the analysis procedure can be optimized, by a proper 
choice of the parameters, such as energy and time windows where to look 
for mass effects.

Concerning the sensitivity to neutrino mass, two points are interesting: the 
minimal detectable $\nut$ mass and the upper limit which can be established on 
$\mnut$ when no positive signal comes out. Correspondingly, we shall address 
the following questions:
\begin{list}{3\alph{enumi}.}{\usecounter{enumi}}
\item What is the minimal $\mnut$ value expected to give a signal 
distinguishable from the statistical fluctuations of the massless case?
\item What is the minimal $\mnut$ value for which the signal expected for 
massless neutrinos can be discriminated from the statistical fluctuations of 
the massive case?
\end{list}

The paper is organized as follows: we develop a method of data analysis, 
based on directional considerations (Section 2) which we use to determine the 
minimal detectable $\nut$ mass for a set of theoretical supernova 
models (Section 3). The same analysis is repeated for a set 
of phenomenological models derived from $SN1987A$ data (Section 4).
We then compare our results with those of ref. \cite{krauss} (Section 5).
Finally, we calculate, both for theoretical and phenomenological models,
the mass upper bound obtainable through the same method (Section 6).
Detector characteristics, cross sections and expression for event rates are 
collected in the Appendix.

\sez{The analysis method: isotropic vs. directional signal.}

Krauss {\it et al.} \cite{krauss} proposed in 1992 a beautiful procedure of 
data analysis for \v{C}erenkov detectors, based on directional considerations.
In this section we briefly recall the idea and then we specify our approach to
this method in terms of: $i)$ the kinematical characterization of neutrino 
events; $ii)$ the assumptions needed to connect the emission features of 
electron and other flavors neutrinos; $iii)$ the quantitative criterion for 
the evidence of a non-vanishing $\nut$ mass.

\subsection{Generalities.}

Let us consider Super-Kamiokande (SK hereafter), a water \v{C}erenkov detector
whose main characteristics are listed in the Appendix, see also \cite{url}.
Neutrino events are originated basically in processes of four kinds:
  
\begin{enumerate}
    \item \it capture on a proton: \rm
\equation \anue + p\stackrel{W}{\longrightarrow} n + e^+ \endequation 
    \item \it scattering off an electron: \rm
\equation \stackrel{\mbox{{\tiny $(\!$}}-\mbox{{\tiny $\! )$}}}{\nue} +e^-
\stackrel{W,Z}{\longrightarrow}\; \stackrel{\mbox{{\tiny $(\!$}}-\mbox{{\tiny 
$\! )$}}}{\nue} + e^- \endequation 
\equation \nui +e^- \stackrel{Z}{\longrightarrow}\;\nui + e^- \endequation
    \item \it capture on an oxygen nucleus: \rm
\equation \nue +\,^{16}\! O \stackrel{W}{\longrightarrow} \,^{16}\! F + e^-
\endequation 
\equation \anue +\,^{16}\! O \stackrel{W}{\longrightarrow}\,^{16}\! N + e^+ 
\endequation 
    \item \it coherent scattering off an oxygen nucleus: \rm
\equation \hskip 60pt \nux +\,^{16}\! O \stackrel{Z}{\longrightarrow} 
\nux +\,^{16}\! O^* \hskip 20pt x=e,\mu ,\tau\endequation 
\end{enumerate}

Expressions for the cross sections are collected in the Appendix. The 
approximate numbers of expected events in SK, for a $SN1987A$-like supernova 
near the galactic center, are shown in Table 1, see also \cite{burr92}.

\begin{table}[htbp] 
\begin{center}
\begin{tabular}{||rcl|c||}
\hline
\hline
 & & & \\[-.8em] 
$\anue p\!\!$ & $\rightarrow$ & $\!\! e^+ n$ & {\bf $\simeq$ 5000} \\ 
$\stackrel{\mbox{{\tiny $(\!$}}-\mbox{{\tiny $\! )$}}}{\nue}e^-\!\!$ & 
$\rightarrow$ & $\!\!\stackrel{\mbox{{\tiny $(\!$}}-\mbox{{\tiny $\! )$}}}
{\nue}e^-$ & {\bf $\simeq$ 100} \\
$\nui e^-\!\!$ & $\rightarrow$ & $\!\!\nui e^-$ & $\bf{\simeq 2 \times 25}$\\
$\stackrel{\mbox{{\tiny $(\!$}}-\mbox{{\tiny $\! )$}}}{\nue}O\!\!$ & 
$\rightarrow$ & $\!\!\stackrel{\mbox{{\tiny $(\!$}}-\mbox{{\tiny $\! )$}}}
{\nue}O$ & {\bf $\simeq$ 50} \\ 
$\nux O\!\!$ & $\rightarrow$ & $\!\!\nux O$ & {\bf $\simeq$ 200} \cite{lang} 
\\[10pt]
\hline
\hline
\end{tabular} 
\cap{Expected events in SK for a supernova at galactic center.}
\end{center}
\end{table}

Among the four processes, scattering off electrons is the only directional
one. For capture reactions and interactions with nuclei, the observable
particles ($e^\pm$ or $\gamma$) are emitted (almost) isotropically, 
whereas in processes
(2.2-3) the recoil electron's angular distribution is strongly forward peaked.
Therefore one can define an appropriate observation angle $\thf$, 
along the supernova direction, which includes all of the electrons from 
$\nu - e$ scattering.

Here is the required signal splitting: outside the forward cone, nearly any
event is originated by $\nue$ and $\anue$. The only contribution of other
flavors  neutrinos is due to (2.6), and can easily be excluded by raising the
minimal detected energy up to 7 -- 8 $\mev$ \cite{lang}. So we can regard 
this signal as untouched by mass effects, and use it to derive the expected 
distribution in the forward cone for the case \mb{\mnut = 0}. 

The analysis procedure is sketched in Figure 1.

\begin{figure}[p]
\centering\leavevmode
\epsfxsize=5.5in\epsffile{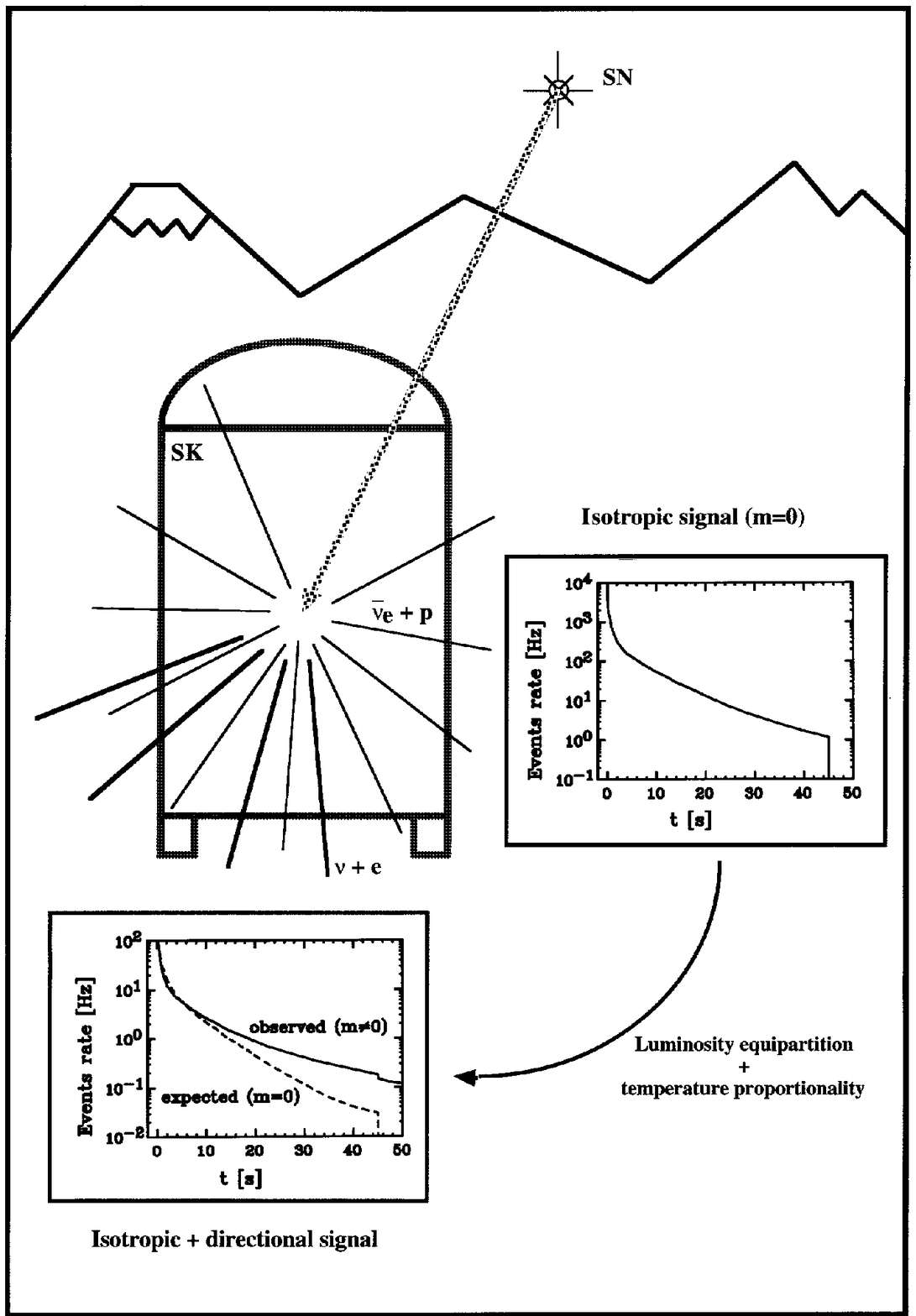}
\cap{Sketch of the directional analysis method. Electron 
(anti-)neutrinos coming from the supernova originate in the detector an
isotropic signal (on the right). This one is used to infer the directional
distribution for massless $\nut$ (dashed line), which is to be compared with 
the observed signal (solid line), see bottom graph.}
\end{figure}

\subsection{Kinematics.}

A specification of the forward observation angle $\thf$ for the analysis.

When a neutrino with energy $\eps$ scatters off an electron at rest which 
gets kinetic energy $\ee$, the electron scattering angle $\vartheta$ is 
determined from energy-momentum conservation:
\equation \cos\, \vartheta  (\eps,\ee) = \sqrt{{ \ee\over \ee + 2m_e}}
\:\left( 1+{m_e\over \eps} \right) \endequation
As $\vartheta$ is a decreasing function of $\ee$ and a rising 
function of $\eps$, the maximum value $\vartheta 
_{\it max}$ is obtained by taking in (2.7) the minimal detected energy $\emin$ 
and taking the limit $\eps\rightarrow \infty$. 
In this way one has:
\equation \vartheta _{\it max} (\emin) = {\rm acos} \sqrt{{ \emin\over \emin 
+ 2m_e}}\endequation
For the lowest value of $\emin$ (i.e. the energy threshold of SK, $\epsilon 
_{th} = 5\:\mev$) this gives \mb{\vartheta _{\it max}\simeq 24^{\circ}}.

In Figure 2 we plot the angular distribution of the $\nut$ scattered electrons 
along the supernova direction, for two significative values of the lower cut: 
$\emin = 5\:\mev$ (SK threshold) and $\emin = 15\:\mev$ (the highest value 
useful for the analysis, as we shall see). Although for these calculations we 
used the ``reference'' model to be described in Section 3, very similar 
distributions are obtained from other supernova models. In the same figure a 
plot of \mb{\vartheta _{\it max}} is given as a function of the lower 
energy cut.

\begin{figure}[p]
\centering\leavevmode
\epsfbox{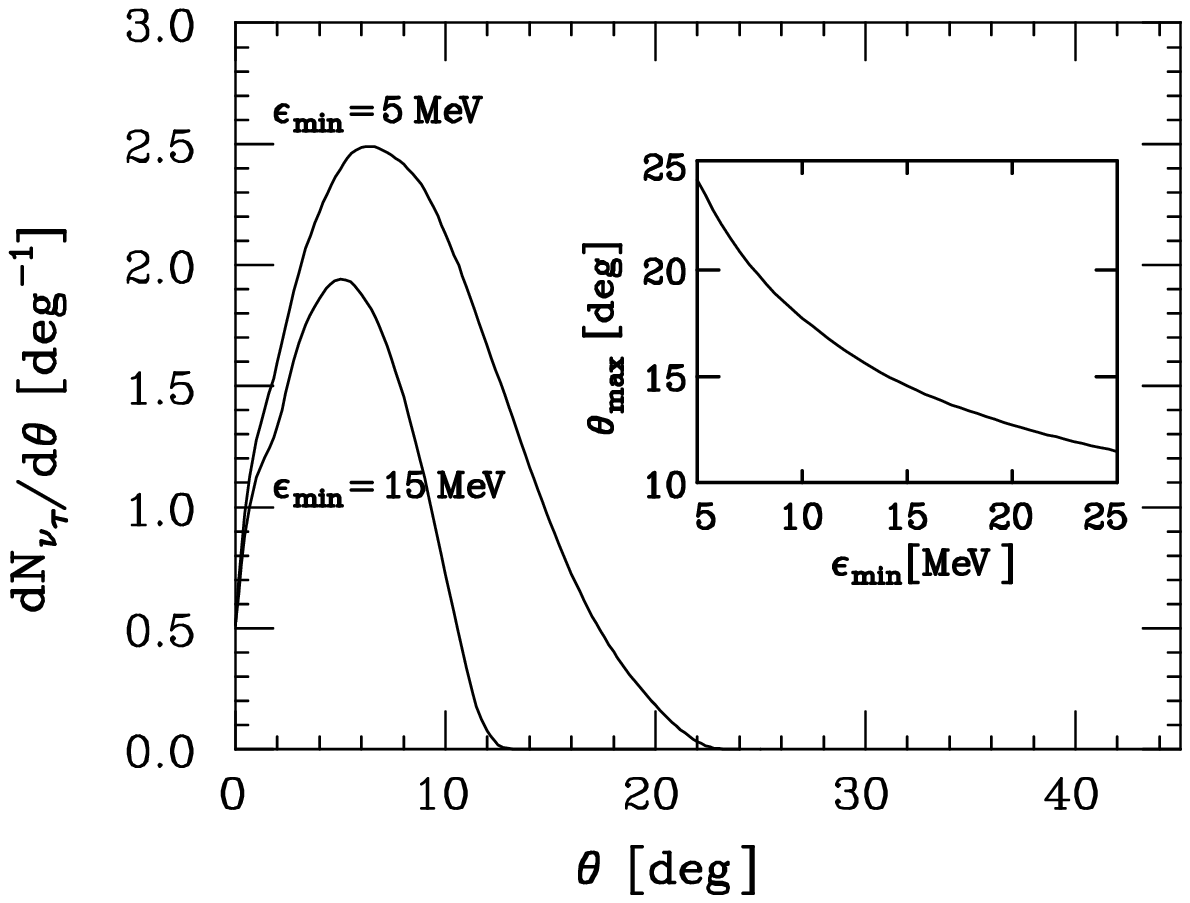}
\cap{The angular distribution of $\nut$ scattered electrons for
two values of $\emin$, and the maximum scattering angle
\mb{\vartheta _{\it max}} as a function of the lower cut. Calculations refer 
to Burrows's reference model, see Section 3.1 for details.}
\end{figure}

\begin{figure}[p]
\centering\leavevmode
\epsfbox{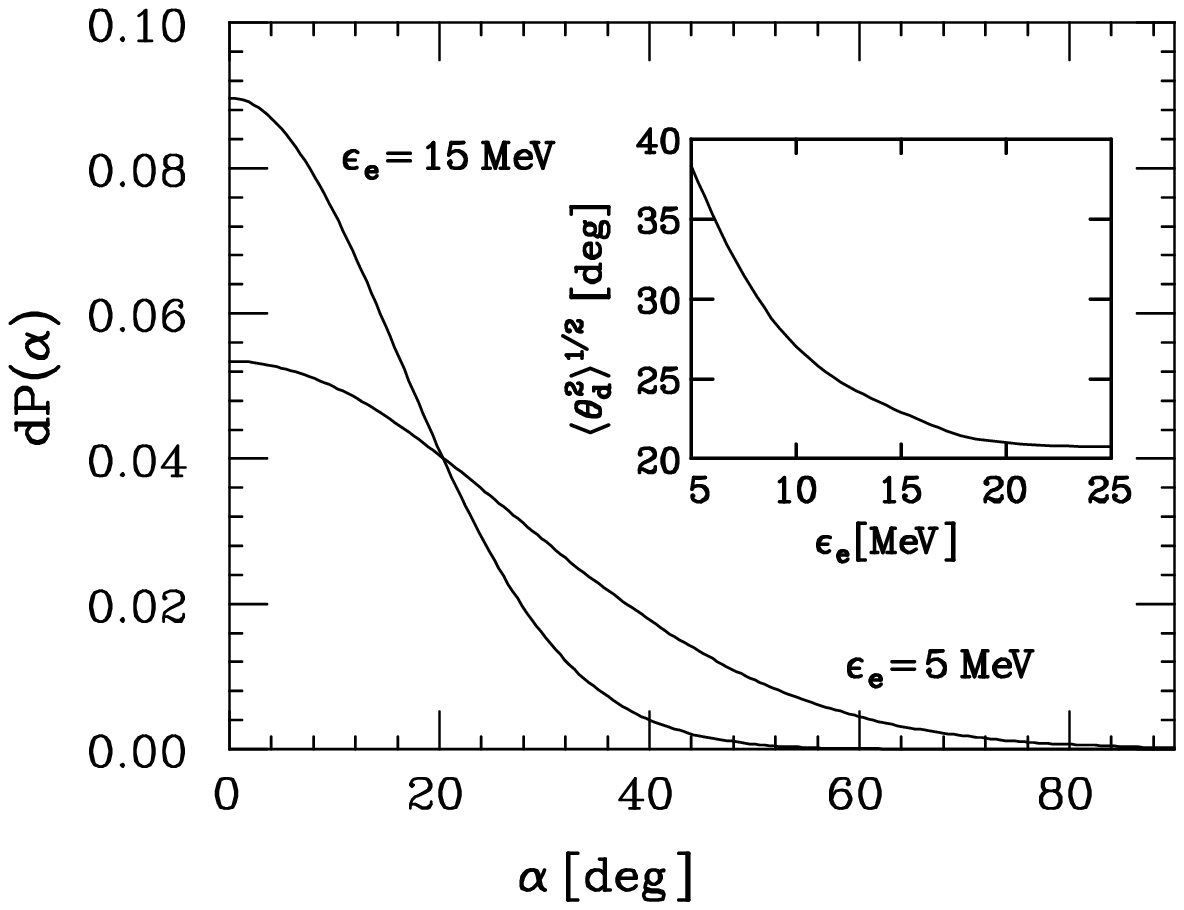}
\caption[The probability to detect a diffused electron at an
angle $\alpha$ from the original trajectory, for two significative values of
$\ee$, and the experimental angular resolution as a function of the electron 
energy in Kamiokande-II.]{\it The probability to detect an electron at an angle
$\alpha$ from the diffusion trajectory, for two values of
$\ee$, and the experimental angular resolution \mb{\langle \vartheta _d ^2 
\rangle ^{1/2}} in Kamiokande-II, from \cite{koshiba}.}
\end{figure}

Besides this purely kinematic dispersion, a systematic angular uncertainty 
is also present, due to the detector. This uncertainty 
is a decreasing function of the scattered electron
energy too, for more energetic particles originate more \v{C}erenkov light, 
and so one can point back to their directions with greater accuracy. 
The probability to detect an electron at an angle $\alpha$ from the 
scattering direction is assumed to be a gaussian function
\equation dP(\alpha,\ee) \propto e^{-{1\over 2}\left( {\alpha \over \sigma 
_\alpha (\ee)}\right) ^2}d\alpha \endequation
where \mb{\sigma _\alpha = {1\over \sqrt{2}} \langle \vartheta _d ^2 \rangle 
^{1/2}} and \mb{\langle \vartheta _d ^2 \rangle ^{1/2}} is the empirically 
determined angular resolution of the detector for an electron with energy $\ee$.

In Figure 3 one finds the the probability to detect an electron at an
angle $\alpha$ from the original trajectory, as given by (2.9), for
$\ee=5,\, 15\:\mev$, and
a plot of the angular resolution dependence on the electron energy 
for the \mbox{Kamiokande-II} detector \cite{koshiba}; a similar 
performance is expected for SK. In the case of a 5 $\mev$ electron one has 
\mb{\langle \vartheta _d ^2 \rangle ^{1/2} \simeq 38^{\circ}}.

Equation (2.9) tells us that one has a non-vanishing 
probability to detect a scattered electron just backward the supernova 
direction! This leads us to formulate a more
reasonable definition for the forward observation cone than ``the cone 
containing {\it all} $\nu + e$ scattering events''.

We verified that the model independent definition
\equation \thf (\emin) \equiv \sqrt{\vartheta ^2 _{\it max}(\emin) + \langle 
\vartheta ^2 _d \rangle(\emin)}\endequation
which can be read as the combination of the kinematic and detector 
spreads just like two independent errors, gives $\thf$ values such that no 
more than one $\nu +e$ event is missed.
For \mb{\emin =5\:\mev} one has \mb{\thf \simeq 45^{\circ}}.
The corresponding solid angle, \mb{\Omega _{fw} = 2\pi\,[1- \cos(\thf)]}, 
for \mb{\emin = 5\: \mev} includes a 95\% of the $\nut$ events, 
but only a 15\% of the total $\anue$ signal.

In Figure 4 one has the angular distribution of the tau-neutrino events, with
respect to the supernova direction, considering both kinematic and detector
spreads, while in Figure 5 we plotted the forward observation angle $\thf 
(\emin)$ as defined in (2.10).

\begin{figure}[p]
\centering\leavevmode
\epsfbox{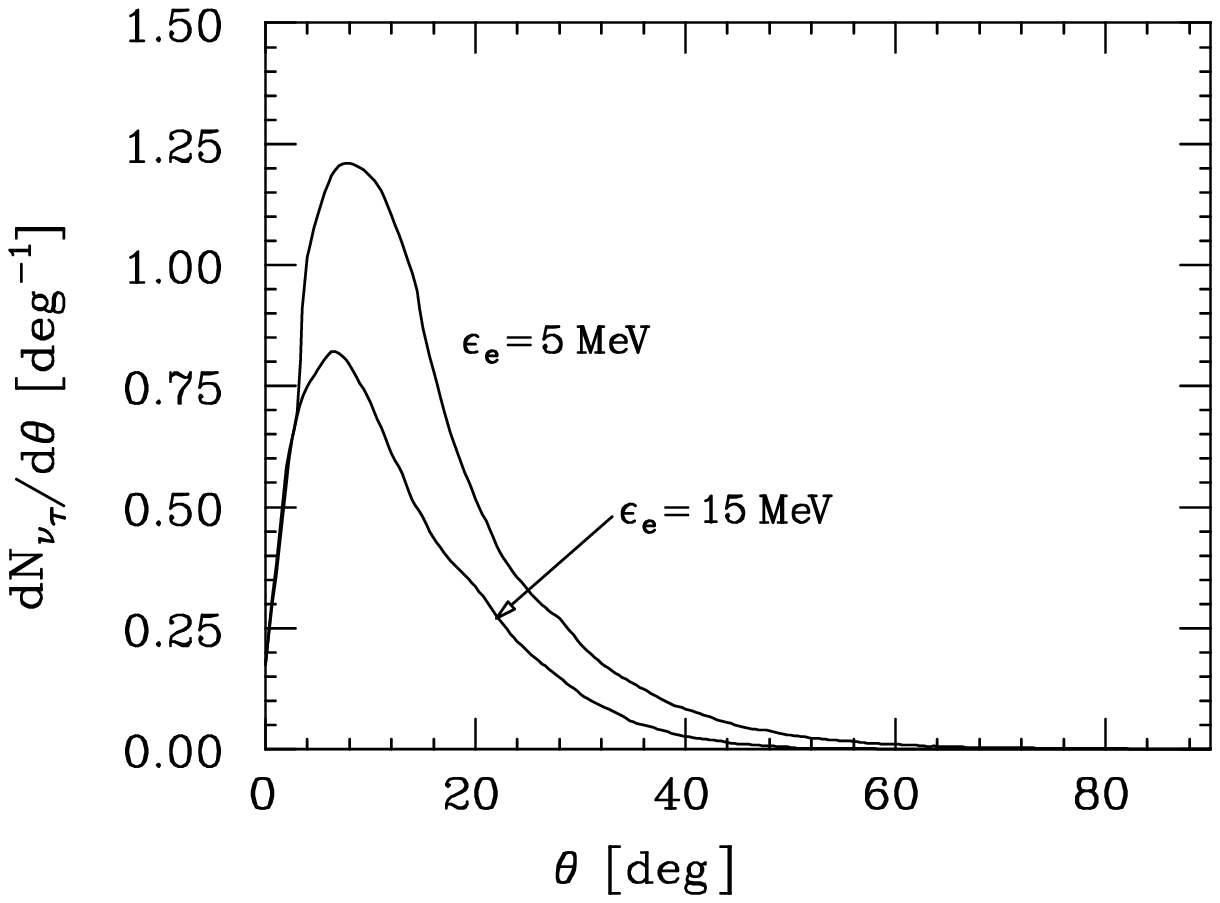}
\cap{The angular distribution of the $\nut +e$ signal with respect to the 
supernova direction, including both kinematic and detector spreads, for two
significative values of $\emin$. Calculations refer to Burrows's reference
model (see Section 3.1).}
\end{figure}

\begin{figure}[p]
\centering\leavevmode
\epsfbox{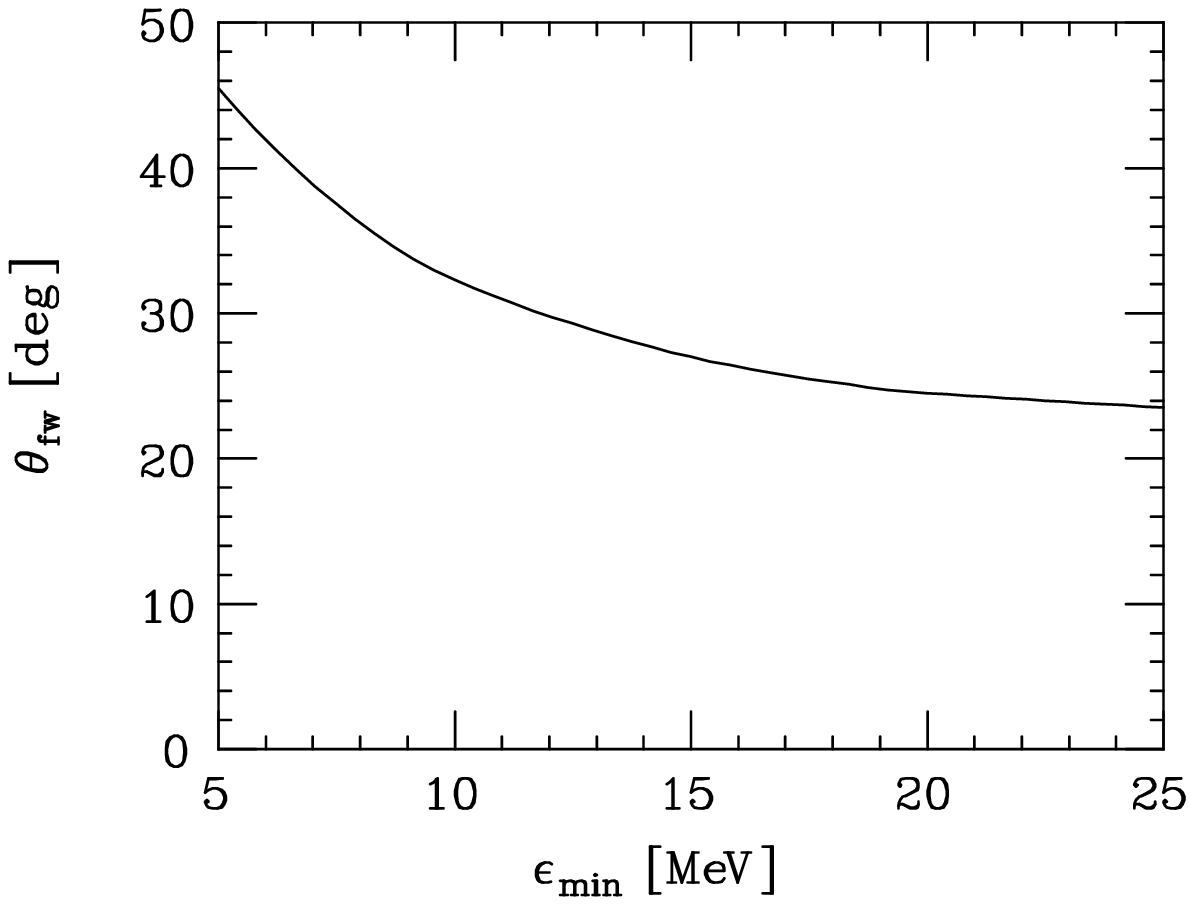}
\cap{The forward observation angle as a function of the lower energy cut.}
\end{figure}

\subsection{Supernova.}

Next, we have to describe how the $\nui$ signal is derived from the $\anue$ 
one.  For this purpose, we use three assumptions, supported by the results of 
several numerical simulations, (see \cite{sim}) which connect the production 
features of electron and other flavor neutrinos {\it in the cooling phase}:

\begin{enumerate}
\item The neutrino energy distributions, at any time and for
each $\nu$ species, are (approximately) thermal, thus specified by the
temperatures of the neutrino-spheres $T_{\nue}(t)$, $T_{\anue}(t)$, 
$T_{\nui}(t)$ and luminosities $L_{\nue} (t)$, $L_{\anue} (t)$,  
$L_{\nui} (t)$. The emitted neutrino rate spectrum \mb{d^2 N_{\nux} / 
(d\eps\, dt)} at time $t$ and energy $\eps$ is therefore:
\equation {d^2 N_{\nux}(\eps ,t) \over d\eps\, dt} = {L_{\nux}(t)\over {\cal F}
_3\: T_{\nux}^4 (t)}\, {\eps ^2 \over 1+e^{\eps / T_{\nux} (t)}}\endequation
where \mb{{\cal F}_3 \equiv \int _0 ^\infty dx\, x^3 (1+e^x )^{-1} \simeq 5.68}
\item Energy equipartition among the six neutrino species holds at 
every time:
\equation L_{\nue} (t) = L_{\anue} (t) = L_{\nui} (t) \endequation
\item The temperatures of the six neutrino species are simply proportional at 
any time:
\equation \left\{ \begin{array}{ccc} T_{\nue} (t) & = & \alpha _{\nue} \cdot 
T_{\anue} (t) \\ T_{\nui} (t) & = & \anut \cdot T_{\anue} (t) \\ 
\end{array} \right. \endequation
\end{enumerate}

Arguments supporting the second assumption can be found in \cite{janka}.
Typical values of $\alpha _{\nue}$ and $\anut$ are:
\equation \begin{array}{ccc} \alpha _{\nue} & = & 0.5 \, -\, 0.9 \\ 
\anut & = & 1.1\, -\, 1.9 \\ \end{array}\endequation
In what follows, we fix the ratio between $\nue$- and $\anue$-sphere 
temperatures at the central value \mb{\alpha _{\nue}=0.7}, while we'll 
keep $\anut$ as a free parameter, in the range given by (2.14).

\subsection{Statistics.}

Finally, we introduce our criterion to estimate the evidence level of
a non-vanishing mass. 

We define \mb{N_m ([t_1,t_2],[\emin,\emax])} the expected number of events in 
the forward observation cone, in the time interval \mb{[t_1,t_2]}, for an 
energy window \mb{[\emin,\emax]}, if the $\nut$ mass is $m$. 

Let us assume that tau-neutrinos have mass $m\neq 0$: we expect the experiment
to detect $N_m$ events. The hypothesis that $\nut$ have a vanishing mass and
that the difference \mb{\abs{N_m - N_0}} is due to a statistical fluctuation
of the massless case can be tested by evaluating the quantity
\equation  s(m,[t_1,t_2],[\emin,\emax]) = { \abs{N_m - N_0} \over 
\sqrt{N_0}} \endequation

Clearly, one can vary \mb{[t_1,t_2]} and \mb{[\emin,\emax]} in order to 
optimize the analysis. The best indicator of a non-vanishing mass is therefore 
the quantity
\equation S(m) = \max _{[t_1,t_2],[\emin,\emax]}\hskip -6pt 
\mbox{{\Large $\{$}} s(m,[t_1,t_2],[\emin,\emax])\mbox{{\Large $\}$}}
\endequation
where the time intervals are chosen arbitrarily within the observation window
(first 100 seconds after bounce), with the only request that at least an
event should be present in \mb{[t_1,t_2]}.
As an example, if \mb{S(100\: \ev) =3} one has a 3$\sigma$ evidence for 
a mass \mb{m=100\: \ev}, i.e. the Confidence Level is 99.7\%.

For brevity, we shall define as ``detectable'' those masses such that 
\mb{S(m) \ge 3}, and we denote the minimal detectable mass as 
\mb{m_{3\sigma}\equiv \min \{ m\, :\, S(m)\ge 3 \}}.

Note that we are using only event counts, disregarding the possible 
additional information of spectral deformations.

A sample of this statistical analysis is reported in Figure 6, where we
plot the expected signal in the forward observation cone for $m=0$ and
$m=200\:\ev$ (for details on the model used for the calculation, see Section 
3). For example, in the time interval \mb{[t_1,t_2]=[25,35]\: s}, and for
\mb{[\emin,\emax]= [15,25]\:\mev}, one expects 1.3 events for the massless 
case, and 4.2 events for massive neutrinos; the evidence index is therefore 
\mb{s(200\ev ,\, [25,35]s,\, [15,25]\mev) = 2.5}. We conclude that, 
in the time and energy intervals  under examination, $m=200\:\ev$ is not 
sufficient to produce a clear effect.

\sez{Results from theoretical models of supernov\ae.}

Till February 1987, numerical simulations were the only way to explore 
supernova neutrinos features. Most of the analyses
just concerned the very initial phase of the emission (\mb{t\lapprox 1\:s}), 
mainly for implications on explosion mechanism,
while only a few dealed with the long term cooling stage, which is relevant for 
us. For a review of simulations see for instance \cite{suz_rev}.

In this section we use the simulations due to Burrows \cite{burr88}, which 
describe the cooling phase, to evaluate the minimal mass $m$ whose effects 
could be observed with the analysis method presented in Section 2. 
We also investigate the dependence of the results on the free
parameters of the analysis.

\subsection{Models and general results.}

We performed a complete set of calculations using several supernova models 
taken from Burrows. In ref. \cite{burr88} seventeen models are presented:
disregarding ``exotics'' and black holes, the remaining ten models look 
consistent with data from $SN1987A$.

Burrows's data, that cover the first 20 seconds for $\anue$ luminosity and 
temperature, were extrapolated to later times by use of power-laws:
$$A(t)=A_0 (1+{t\over \tau _A})^{-n_A}$$
where $A$ denotes here both $L_{\anue}$ and $T_{\anue}$, and the parameters
$A_0$, $\tau _A$ and $n_A$ are obtained from fits to late times behaviours
(\mb{t>10\: s}). We also introduced a time parameter $\tt$ beyond which 
the neutrino luminosities vanish: this happens when the 
proto-neutron star becomes transparent to neutrinos. Typical values for
$\tt$ are (40 -- 50) $s$ \cite{burr92}. Finally, for the other neutrino
species, luminosities and temperatures are determined through relations 
(2.12-13). 
Unless otherwise stated, all results reported in the text, tables and figures, 
refer to the Super-Kamiokande detector for the following ``default'' set of 
parameters: 
\equation \begin{array}{lcll} \anut & = & 1.5 & \\ \tt & = & 45 & s \\ 
D & = & 10 & kpc \\ \end{array} \endequation 

For each model, the main characteristics  are listed in Table 2, together with 
the results:  the optimal choices for time and energy windows, 
and the value of the minimal detectable mass, as defined in Section 2.4.
The $\anue$ event rate and the sensitivity to non-vanishing neutrino mass are 
shown in Figures 7 and 8 for a few representative models: the {\it M}ost
{\it F}avourable case (\# 52), the {\it L}east {\it F}avourable case (\# 62) 
and the {\it RE}ference model (\# 55), which very well accounts for $SN1987A$ 
data.

We notice that the minimal detectable mass lies anyhow outside the 
cosmologically interesting region: even in the most favourable model
we have \mb{m_{3\sigma} \simeq 120\: \ev}. For the reference
model one has \mb{m_{3\sigma} \simeq 140\: \ev}.

When looking at the dependence on the model characteristics, we find that 
in general: $i)$ for a given equation of state (EOS), mass effects are least 
pronounced for more massive cores; $ii)$ for a given core mass, neutrino mass 
effects are weaker for the soft EOS.

Those features can be understood, at least qualitatively. We remark that, for a 
given EOS, the star radius decreases with mass, so that in more massive cores
matter is more compressed. Also, for a given mass, softer EOS allow higher 
compression. Thus higher masses and/or softer EOS imply higher densities, and 
consequentely higher temperatures and opacities. More energetic neutrinos are 
produced, so that $\nut$-mass induced delays are shorter. Furthermore, and more 
important, the longer cooling phase results in a higher background $\anue$ 
flux at later times.
The increase of the cooling time due to mass and EOS is shown in Figure 7: 
MF and RE models have the same EOS but different masses, whereas LF and RE 
models have the same mass but different EOS.

\begin{figure}[p]
\centering\leavevmode
\epsfbox{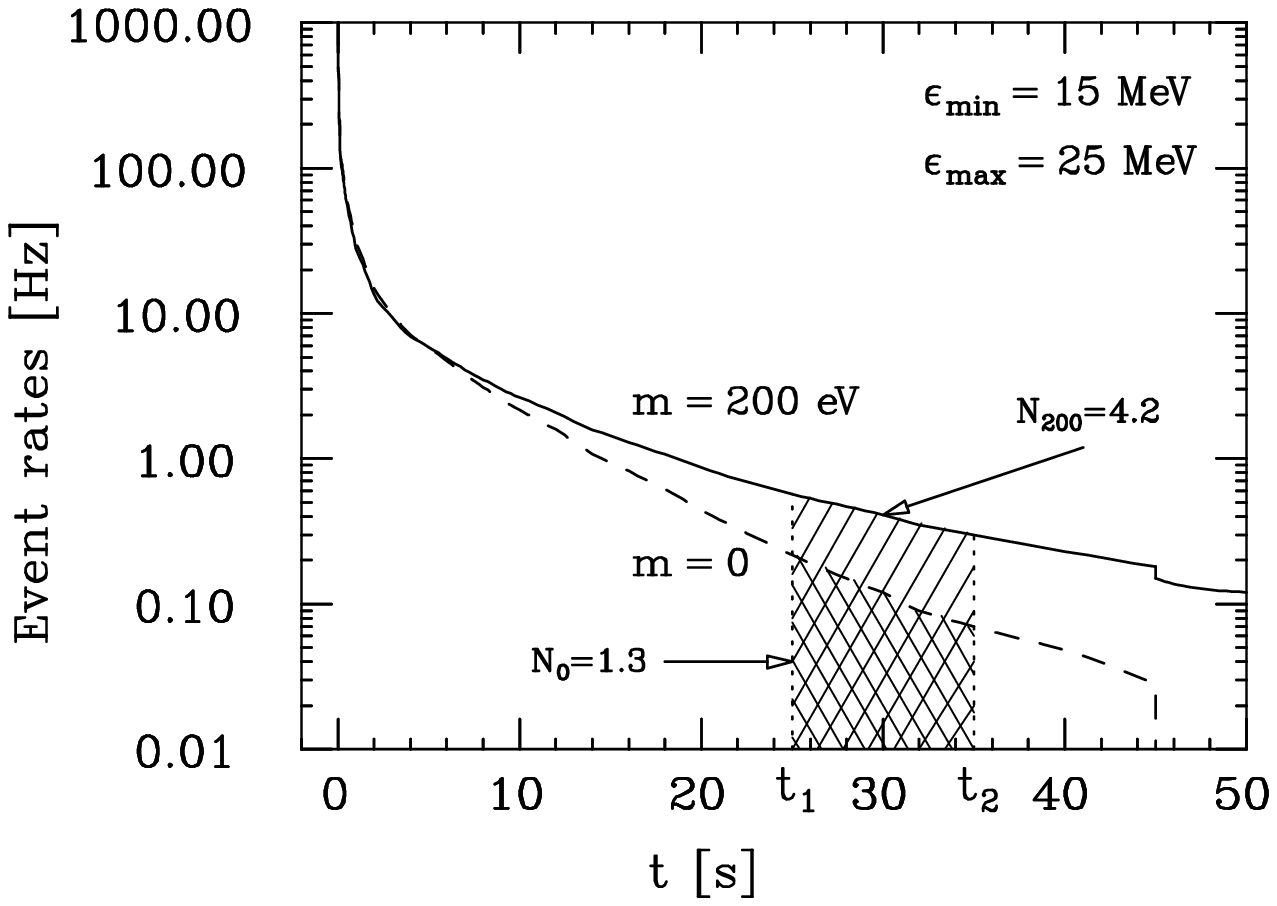}
\cap{An example of the statistical analysis: the dashed line refers to the 
expected distribution in the forward observation cone for the case $m=0$, while
the solid one is for $m=200\:\ev$, see text for explanations. Burrows's 
``reference'' model described in Section 3 was used for this example.}
\end{figure}

\begin{table}[p] 
\begin{center}  
\begin{tabular}{||c||c|c|c|c||c|c|c||}  
\hline  
\hline 
 & & & & & & & \\[-.8em] 
Model & EOS & $M_i\:[M _\odot]$ &  $M_f\: [M _\odot]$ & $E_{tot}$ [{\it foe}] &
$[\emin,\emax]$ [$\mev$] & $[t_1,t_2]$ [$s$] & $m_{3\sigma}$ [\ev] \\[.4em] 
\hline
\hline
52 & stiff & 1.2 & 1.2 & 152 & [10,20] & [25,85] & 120 \\
\hline 
53 & stiff & 1.3 & 1.3 & 171 & [10,20] & [30,95] & 130 \\
\hline
54 & stiff & 1.4 & 1.4 & 191 & [10,20] & [30,100] & 130 \\
\hline
\bf 55 &\bf stiff &\bf 1.3 & 1.5 &\bf 228 &\bf [10,20] &\bf [40,95] &\bf 140 \\
\hline
56 & stiff & 1.3 & 1.6 & 262 & [10,20] & [45,100] & 130 \\
\hline
57 & stiff & 1.3 & 1.8 & 326 & [10,20] & [40,95] & 130 \\
\hline
59 & soft & 1.2 & 1.2 & 158 & [10,20] & [35,100] & 140 \\
\hline
60 & soft & 1.3 & 1.3 & 178 & [10,20] & [40,100] & 140 \\
\hline
61 & soft & 1.3 & 1.4 & 205 & [10,20] & [40,100] & 140 \\
\hline
62 & soft & 1.3 & 1.5 & 235 & [10,20] & [40,100] & 140 \\
\hline
\hline 
\end{tabular} 
\caption[Characteristics and results of Burrows's supernova models.]
{\it Characteristics and results for Burrows's supernova models: the columns 
indicate the equation of state (EOS), the progenitor core mass $M_i$,
the proto-neutron star mass $M_f$ (after accretion), the total emitted 
energy $E_{tot}$ (1 foe = \mb{10^{51}\: erg}), the optimal energy 
$[\emin,\emax]$ and time $[t_1,t_2]$ windows in order to probe mass effects, 
and the minimal detectable mass $m_{3\sigma}$.}
\end{center} 
\end{table} 

\begin{figure}[p]
\centering\leavevmode
\epsfbox{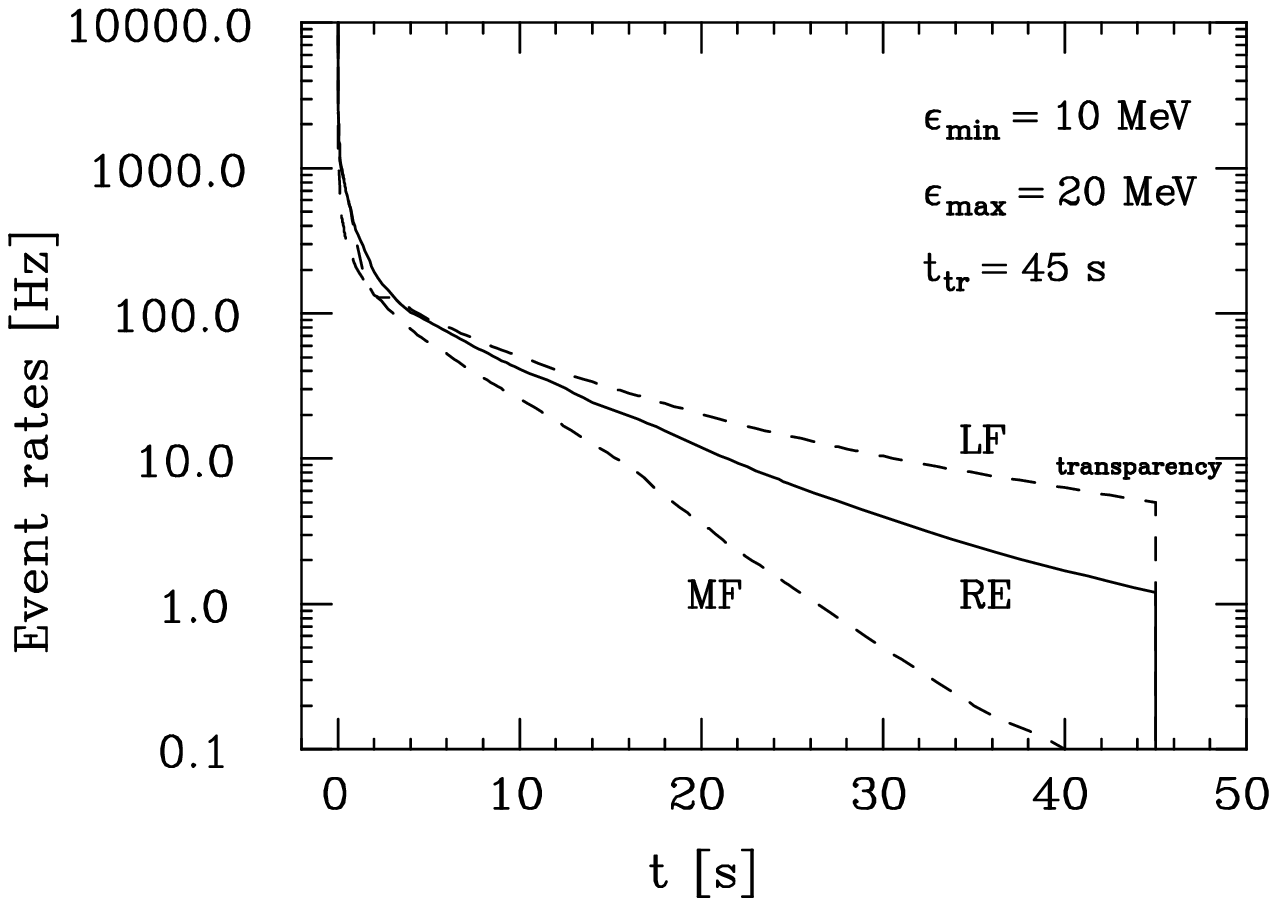}
\cap{The isotropic $\anue$ signal in the whole SK detector, for the least/most 
favourable model (LF/MF), and for the ``reference'' model (RE), see text.}
\end{figure}

\begin{figure}[p]
\centering\leavevmode
\epsfbox{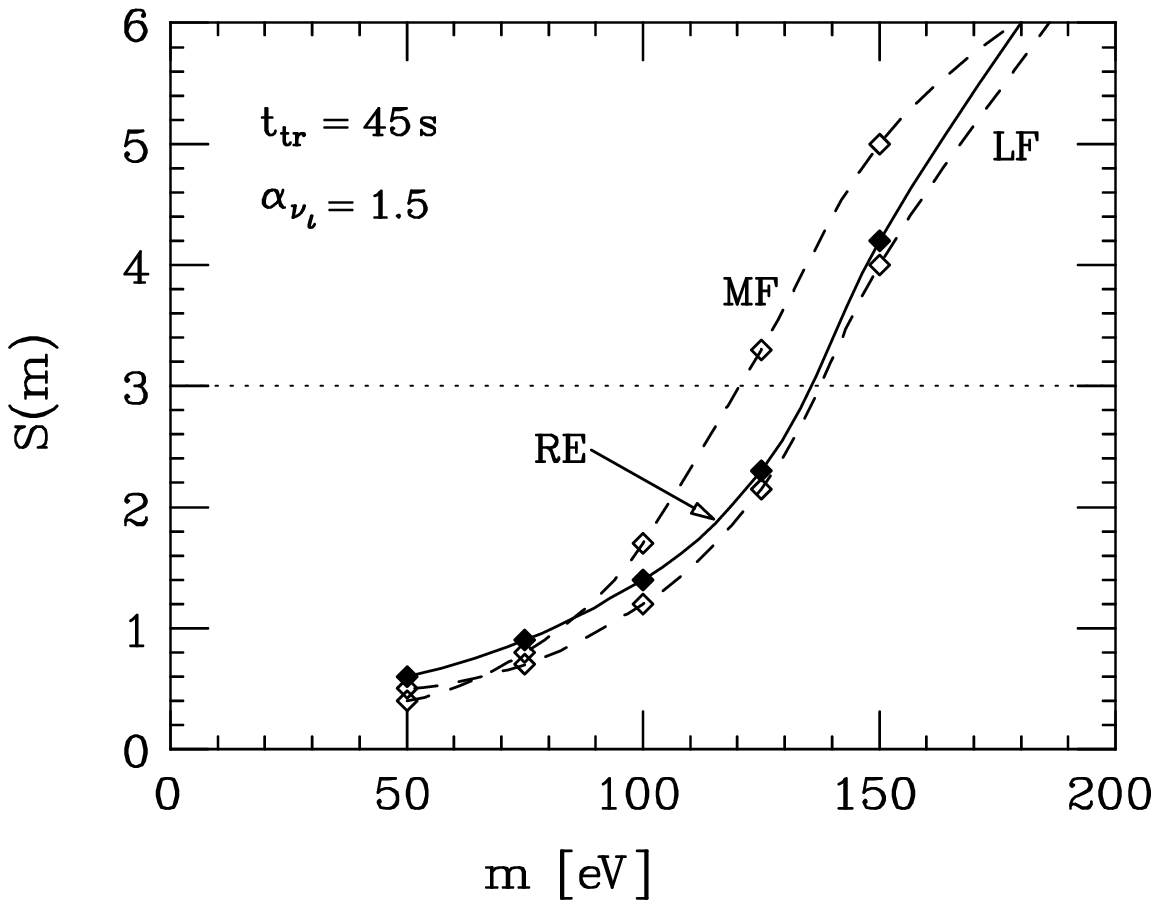}
\cap{Mass sensitivity for a few Burrows's models: same notations as in 
Figure 7.}
\end{figure}

\subsection{Dependence on the model parameters.}

In order to investigate the dependence of the previous results on the
emission parameters, we repeated our calculations by varying one 
parameter at a time, while keeping all the others constant at the values of 
(3.1). The results quoted in the following are for the reference model, 
but we obtained similar behaviours for the others.

We first varied the parameter $\anut$, defined in Section 2.3 as the ratio 
of $\nui$-- to $\anue$--sphere temperatures, repeating the analysis 
for five different values in the range given by (2.14).

The results of this search are summarized in Table 3, where we report, for
each value of the parameter, the minimal detectable mass, together with the
number of $\nut$ events and the characteristic delay time for \mb{m=150\:\ev}. 
In Figure 9 we plot the evidence index $S(m)$ for the central value 
and the two extrema of the $\anut$ range.
The minimal detectable mass depends rather 
weakly on the $\nui$-sphere temperature. The point is that when varying
$\anut$ two competing effects arise. For example, when $\anut$ is increased:
$i)$ the $\nut$-sphere temperature is higher, leading to more energetic
neutrinos, and so to shorter delays;
$ii)$ the number of $\nut$ events increases, leading to a higher statistics.
Our calculations essentially show that these effects almost balance each other,
so that small variations of $\anut$ do not affect substantially the
sensitivity to a non-vanishing mass.

\begin{figure}[p]
\centering\leavevmode
\epsfbox{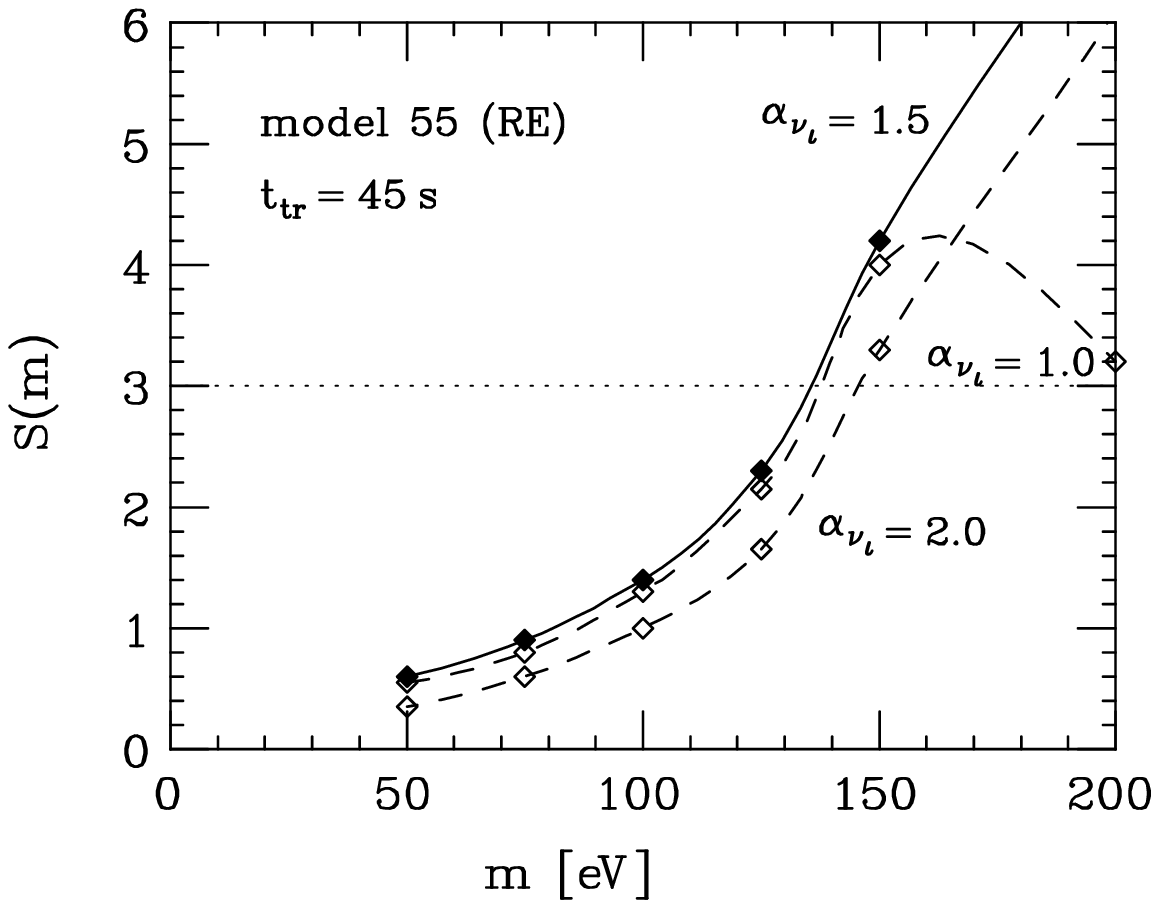}
\cap{Mass sensitivity for several values of the $\nui$-sphere temperature, in 
Burrows's reference model.}
\end{figure}

\begin{table}[p] 
\begin{center}  
\begin{tabular}{||c|c|c||c|c|c||}  
\hline  
\hline 
 & & & & & \\[-.8em] 
$\anut$ & $N_{\nut}$ & $\Delta t (\eps =\anut\!\cdot\! 15 \mev)$ [{\it s}] & 
$[\emin,\emax]$ [$\mev$] & $[t_1,t_2]$ [$s$] & $m_{3\sigma}$ [\ev] \\[.4em]
\hline
\hline
1.0 & 4.93 & 52 & [10,20] & [35,95] & 140 \\
1.3 & 11.8 & 31 & [10,20] & [35,100] & 140 \\
\bf 1.5 & \bf 16.4 & \bf 23 & \bf [10,20] & \bf [40,100] & \bf 140 \\
1.7 & 16.8 & 18 & [10,20] & [40,100] & 140 \\
2.0 & 17.5 & 13 & [10,20] & [40,100] & 150\\
\hline
\hline
\end{tabular}
\cap{Mass sensitivity dependence on the $\nui$-sphere temperature in 
Burrows's reference model, with  $\tt = 45\: s$. $N_{\nut}$ is 
the number of $\nut$ events in the forward cone and $\Delta t$ is the 
characteristic delay time for a neutrino in the explorable mass range 
($m=150\:\ev$) and for $D=10\, kpc$.}
\end{center}
\end{table}

Concerning the dependence on the transparency time,
we performed three calculations of the minimal detectable mass for $\tt$
around the value given by \cite{burr92}. Again the results are 
essentially stable, see Figure 10, where we plotted the evidence index $S(m)$ 
for the inquired values.
Clearly, the more sudden the transparency occurs, the more evident the
delayed neutrinos will be: should we take the (maybe) unphysical
choice \mb{\tt =30\: s}, we find \mb{m_{3\sigma}\simeq 110\:\ev}.

\begin{figure}[p]
\centering\leavevmode
\epsfbox{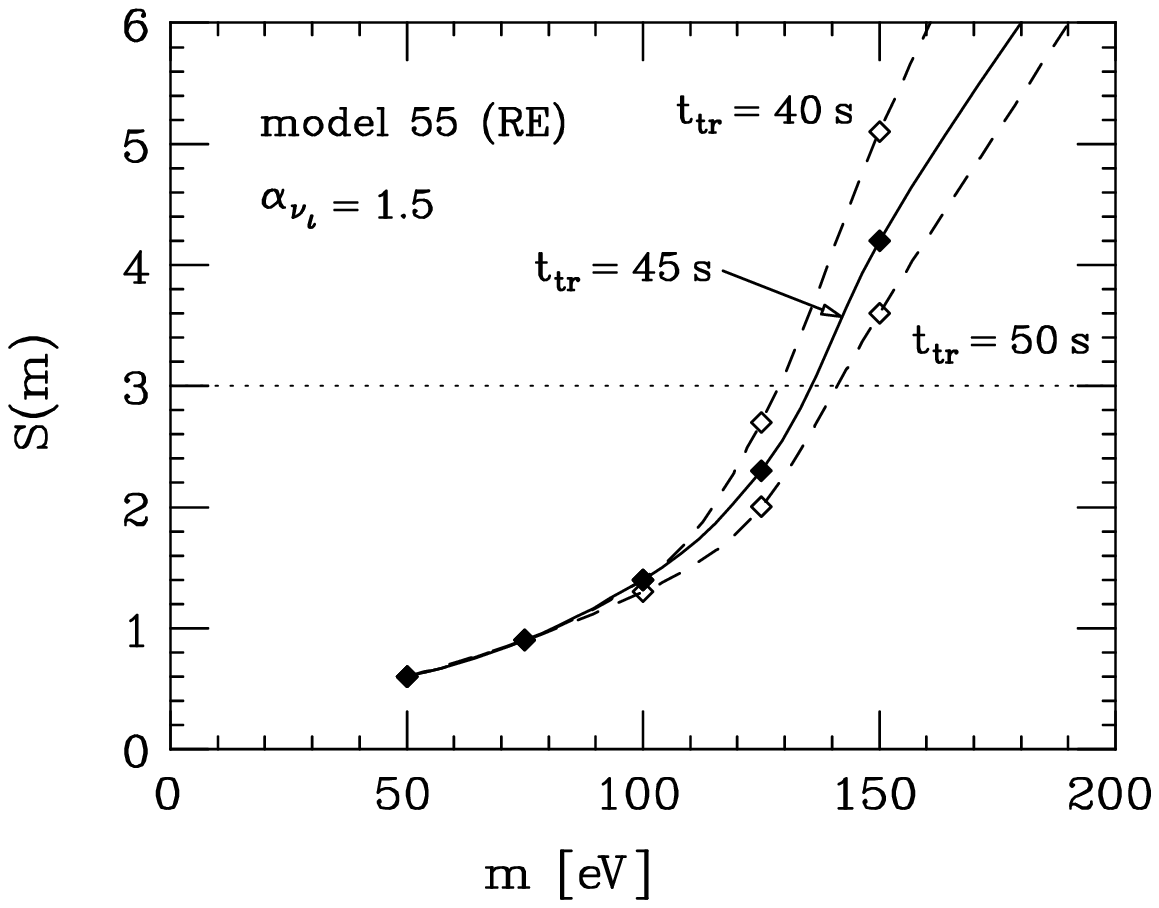}
\cap{Mass sensitivity for several values of the transparency time, in Burrows's 
reference model.}
\end{figure}

\begin{table}[p] 
\begin{center}  
\begin{tabular}{||c|c||c|c|c||}  
\hline  
\hline 
 & & & & \\[-.8em] 
$\tt$ [{\it s}] & $N_{\nut}(\tt\rightarrow t_{\it max})$ & 
$[\emin,\emax]$ [$\mev$] & $[t_1,t_2]$ [$s$] & $m_{3\sigma}$ [\ev] \\[.4em]
\hline
\hline
30 & 6.2 & [10,20] & [30,70] & 110 \\
\hline
40 & 3.9 & [10,20] & [35,95] & 130 \\
\bf 45 & \bf 3.2 & \bf [10,20] & \bf [40,100] & \bf 140 \\
50 & 2.5 & [10,20] & [40,100] & 140 \\
\hline
\hline
\end{tabular}
\cap{Mass sensitivity dependence on the transparency time, in Burrows's
reference model, with $\anut = 1.5$.
\mb{N_{\nut}(\tt\rightarrow t_{\it max})} is the number of $\nut$ events 
expected to occur after the transparency, in the case of $m=150\:\ev$ and for
$D=10\, kpc$.}
\end{center}
\end{table}

Finally,....

\begin{figure}[p]
\centering\leavevmode
\epsfbox{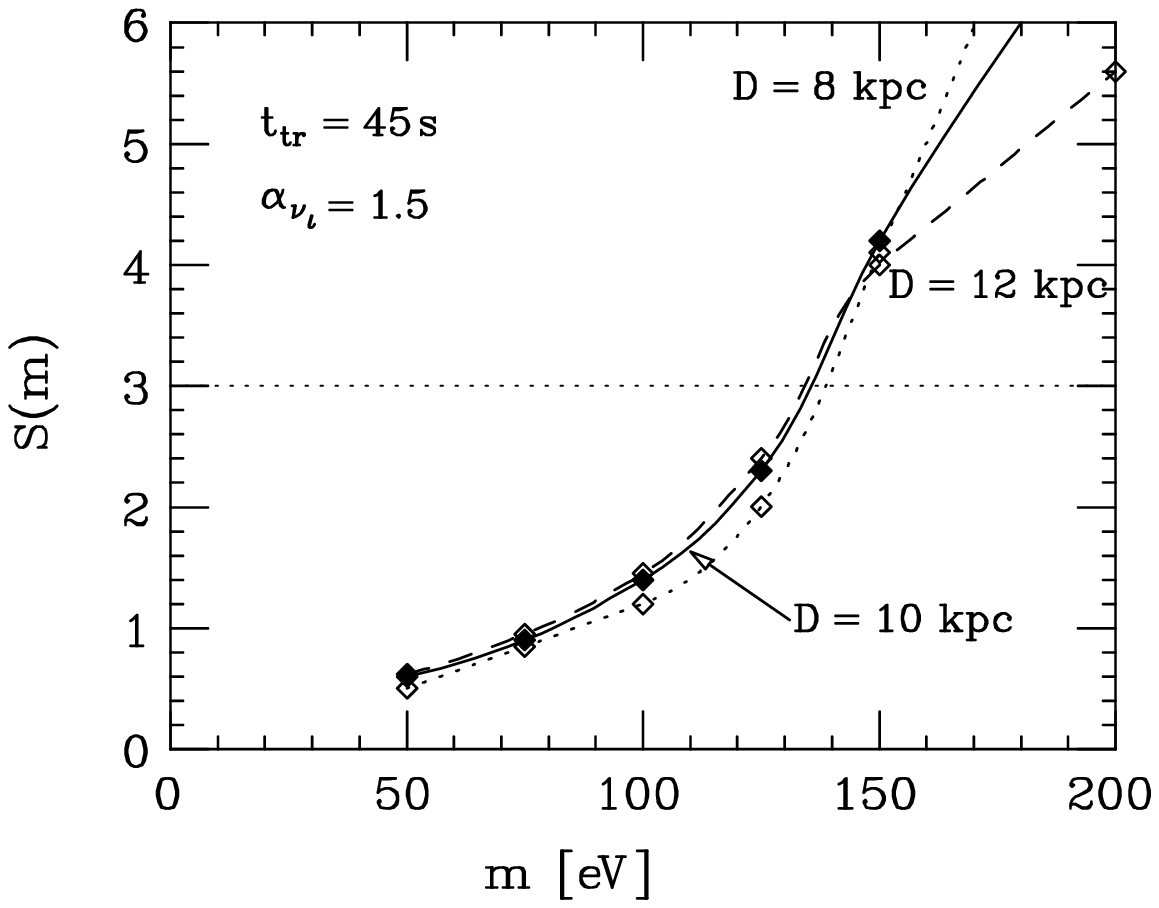}
\cap{Mass sensitivity for several values of the supernova distance, in 
Burrows's reference model.}
\end{figure}

\begin{table}[p] 
\begin{center}  
\begin{tabular}{||c|c|c||c|c|c||}  
\hline  
\hline 
 & & & & & \\[-.8em] 
$D$ [{\it kpc}] & $N_{\nut}$ & $\Delta t (\eps =20\mev)$ [{\it s}] &
$[\emin,\emax]$ [$\mev$] & $[t_1,t_2]$ [$s$] & $m_{3\sigma}$ [\ev] \\[.4em]
\hline
\hline
8 & 25.5 & 14.4 & [10,20] & [40,100] & 140 \\
\bf 10 & \bf 16.4 &\bf 18.0 & \bf [10,20] & \bf [40,100] & \bf 140 \\
12 & 11.4 & 21.6 & [10,20] & [35,100] & 130 \\
\hline
50 & 1.3 & 90.0 & [5,100] & -- & -- \\
\hline
\hline
\end{tabular}
\cap{Mass sensitivity dependence on the supernova distance, in Burrows's
reference model, with $\anut = 1.5$ and $\tt=45\, s$.
$N_{\nut}$ is the number of $\nut$ events and $\Delta t$ is the typical 
neutrino time-delay in the case of $m=150\:\ev$.}
\end{center}
\end{table}

\subsection{Best choices for the analysis parameters.}

These are particularly important dependences to investigate, because we deal
now with detection parameters, which we can change in order to optimize the 
analysis.

It was already noted \cite{krauss} that an upper energy cut enhances the 
mass sensitivity of the analysis method. This is due to the fact that the 
isotropic signal occurs at a higher energy with respect to the directional one. 
In fact in the dominant isotropic process, i.e. $\anue$ capture on a proton, 
the emitted positron takes almost all of the neutrino energy, while in the 
$\nut$ process, namely scattering off an electron, the average energy of the 
outgoing particle is roughly $\ee \sim {1\over 2}\eps$. 
We stress that by means of $\emax$ we cut high energy $\nut$ events, which have
short mass-induced delays, so that the loss in sensitivity is low.
We should be aware, anyway, that this dynamical effect is partially balanced 
by the higher $\nut$ energy at the emission.

Concerning the lower energy cut that we introduced,
many competing effects are present, leading to a more complex situation. 
Basically, at high values of $\emin$ the number of
$\nut$ events get small and, most of all, we deal only with
high energy neutrinos, for which the flight-time delay is shorter.
On the other hand, as we said in Section 2.2, the forward 
observation angle is a decreasing function of the lower cut (see eq. 2.10).
Rising $\emin$ leads therefore to a narrower cone, in which the ratio
between directional and isotropic $\anue$ signal becomes more favourable.

In Table 5 we report the full results of our analysis on the reference model 
for six different energy windows, while in Figure 11 we plot the mass 
sensitivity for three significative choices. 

As we can see from Figure 11, the best choice depends on the explored mass 
region, because of the $m$ dependence in the delay formula (1.1): the larger
is $m$, the higher should be the lower cut. For our
explorable range \mb{m=100 - 200\: \ev} the choice \mb{\emin =10\:\mev;\: 
\emax =20\:\mev} gives the best sensitivity for all models and for 
all values of the emission parameters $\anut$ and $\tt$.

With regard to the time window, as we can see from Table 2 the best choice 
again depends on the explored mass: larger $m$ mean more delayed 
neutrinos, which are to be observed at later times. For our explorable range
the choice \mb{t_1 =40\: s;\: t_2 =100\: s} seems in general the most 
appropriate.

\begin{figure}[p]
\centering\leavevmode
\epsfbox{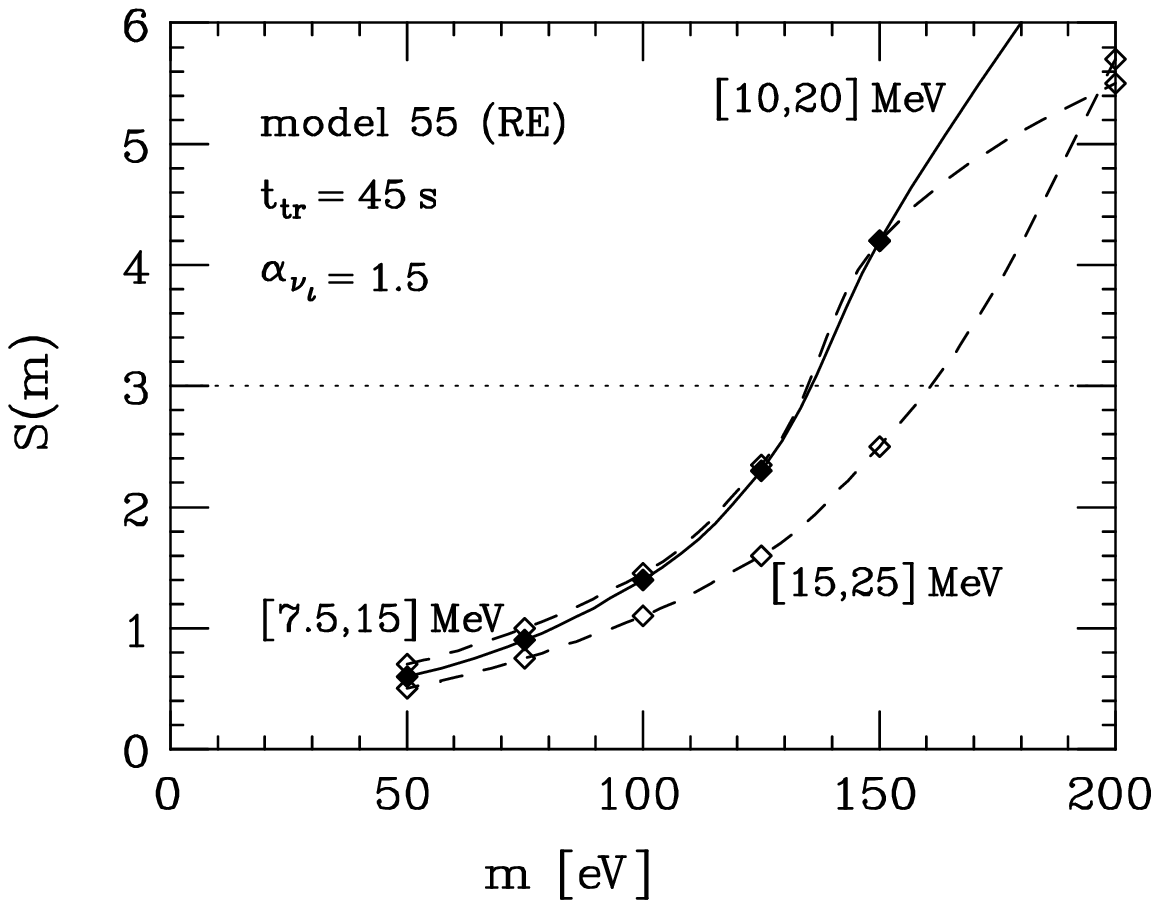}
\cap{Mass sensitivity for significative choices of the energy window 
\mb{[\emin,\emax]}, in Burrows's reference model.}
\end{figure}

\begin{table}[p] 
\begin{center}  
\begin{tabular}{||c|c|c|c|c|c||c||}  
\hline  
\hline 
 & & & & & & \\[-.8em] 
$[\emin,\emax]$ & $N_{\nut} / N_{tot}$ [\%] & $\thf$ & $N_{\nut}/N_{tot}$ [\%] &
$\Delta t _{\it max}$ [{\it s}] & $[t_1,t_2]$  & $m_{3\sigma}$ \\

$[\mev]$ & (all SK) & $[{\it deg}]$ & (in $\thf$) & 
$\equiv \Delta t (\eps =\emin)$ & [$s$] & [\ev] \\[.4em] 
\hline
\hline
$[7.5,15]$ & 0.91 & 37 & 6.4 & 210 & [40,100] & 140 \\
\hline
$\bf [10,20]$ & \bf 0.66 & \raisebox{-10pt}{\bf 32} & \bf 6.5 & 
\raisebox{-10pt}{\bf 120} & \bf [40,100] & \bf 140 \\
$[10,100]$ & 0.52 & & 5.4 & & [30,100] & 190 \\
\hline
$[15,25]$ & 0.48 & \raisebox{-10pt}{27} & 7.0 & \raisebox{-10pt}{52} & [20,80] 
& 160 \\
$[15,100]$ & 0.43 & & 6.4 & & [20,90] & 170 \\
\hline
$[20,100]$ & 0.39 & 24 & 6.8 & 29 & [20,90] & 180 \\
\hline
\hline
\end{tabular}
\cap{Mass sensitivity for significative energy windows, in Burrows's
reference model. Calculations are for $\tt = 45\: s$ and $\anut = 1.5$.
\mb{\Delta t _{\it max}} is the maximum delay for a neutrino in the 
explorable mass range ($m=150\:\ev$), corresponding to the case of minimal 
detected energy, for $D=10\, kpc$.}
\end{center}
\end{table}

\subsection{This section's conclusions.}

We conclude, for the case of a Burrows-like supernova near the galactic center,
that: 
\begin{enumerate}
\item there is no way of getting evidence of \mb{\mnut \lapprox 120 \,\ev};
\item this conclusion is essentially independent of the value of the uncertain
emission parameters $\anut$ and $\tt$;
\item for the explorable mass range, the optimal energy window is 
\mb{\ee \sim [10,20]\: \mev}.
\item for the explorable mass range, the optimal time window is 
\mb{t\sim [40,100]\: s}.
\end{enumerate}

\sez{Results from phenomenological models of supernov\ae.} 

So far we considered theoretical models for neutrino emissions, essentially
based on numerical simulations.
In order to investigate further the potentialities of the analysis method, 
in this section we consider phenomenological models, built so as to reproduce 
the data from $SN1987A$.

The mere 19 events of the 1987 supernova were not sufficient to allow a 
detailed reconstruction of the emission features, and several models were
obtained by fitting the data with different analytical functions. We selected
two of these models, belonging to significative classes, to explore a
wider range of possibilities.

\begin{figure}[p]
\centering\leavevmode
\epsfbox{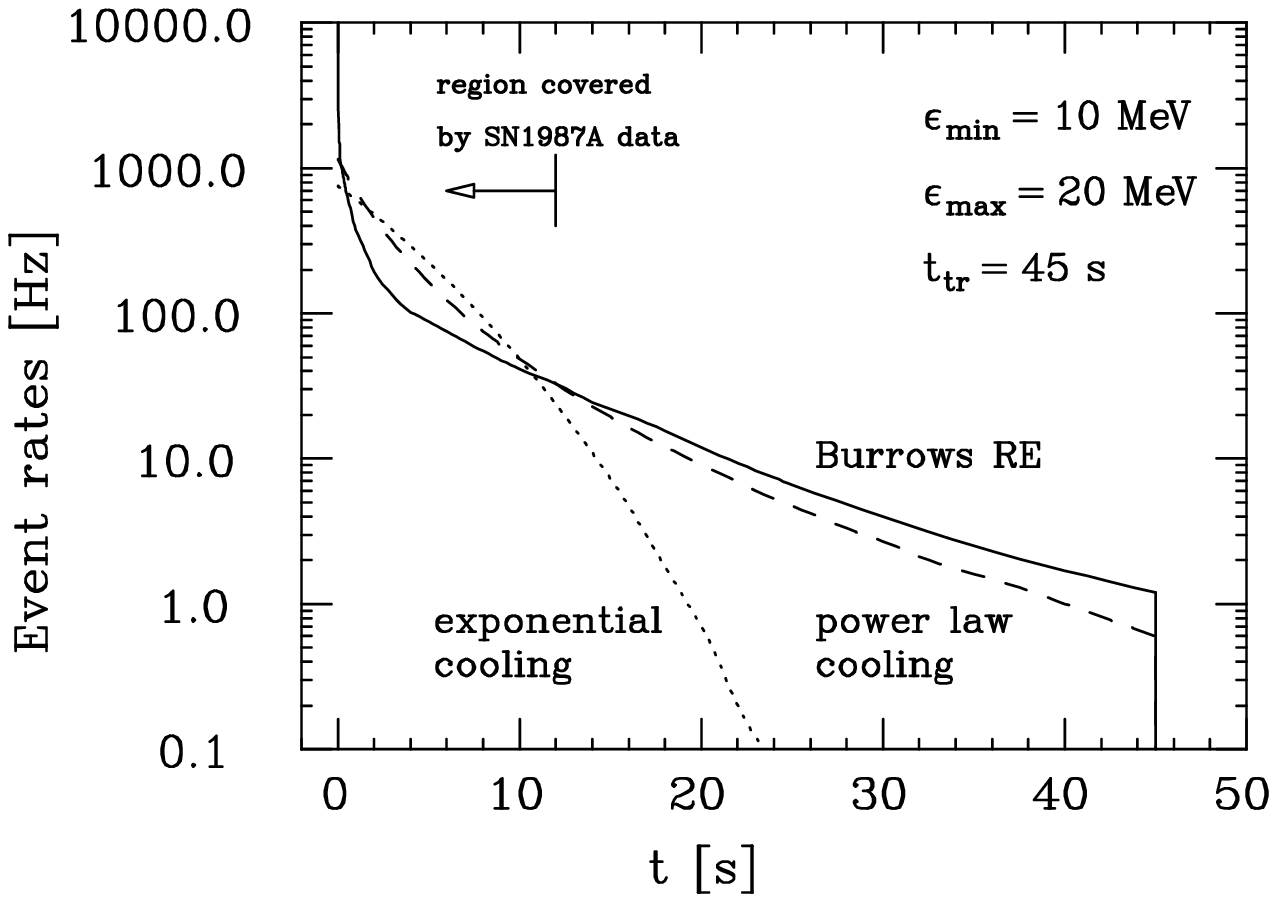}
\cap{Total signals in the whole detector for exponential cooling
(dotted), power-law cooling (dashed), and Burrows's reference model (solid).}
\end{figure}

\subsection{Power-law cooling.}

In this framework, we first consider a power-law parametrization,
defined by the relations:
\equation \left\{ \begin{array}{lll} T(t) & = & T^0 \,\left( 1+{t\over \tau} 
\right) ^{-n}\\ L(t) & = & L^0 \,\left( 1+ {t\over \tau} \right) ^{-4n} \\ 
\end{array} \right. \endequation
Note that, being \mb{L(t)\propto T^4 (t)}, in this model the radii of the 
neutrinospheres are assumed as constant.

We use as best fit parameters for $SN1987A$ those obtained by Bludman \& 
Schinder \cite{b&s}:
\equation  \left\{  \begin{array}{llll}  T_{\anue}^0 & \simeq & 4.20 & \mev \\
L_{\anue}^0 & \simeq & 19.8 & foe/s \\ \tau & \simeq & 2.78 & s \\ 
n  & \simeq & 0.4 &  \\ \end{array} \right.  \endequation
Again we used relations (2.12-13) to derive luminosities and temperatures
for the other neutrinos species.

As we can see from Figure 12, where we plotted the $\anue$ event rates in SK 
for (4.1) and for Burrows's reference 
model, the behaviour of the power-law cooling is similar to that of
theoretical models of Section 3.

With the values of (4.2) and for the default set of parameters (3.1), 
we obtained the minimal detectable mass \mb{m_{3\sigma}\simeq 110\:\ev}. 
This result is consistent with what found for Burrows theoretical models, 
and is again outside the cosmologically interesting range.


\subsection{Exponential cooling.}

A simpler analytic parametrization is obtained assuming that the neutrinosphere
temperatures fall exponentially, while the radii are again kept constant.
One has this way
\equation \left\{ \begin{array}{lll} T (t) & = & T^0 \, e^{-t/4\tau} \\  
L(t) & = & L^0 \, e^{-t/\tau}\\ \end{array} \right. \endequation

The best fit parameters for $SN1987A$ are taken from Loredo \& Lamb
\cite{l&l}:
\equation \left\{ \begin{array}{llll} T_{\anue}^0 & \simeq & 4.16 & \mev
\\ L_{\anue}^0 & \simeq & 12.9 & foe/s \\ \tau & \simeq & 4.6 & 
s \\ \end{array} \right. \endequation

With these choices one obtains a minimal detectable mass \mb{m_{3\sigma}\simeq 
70\:\ev}.
The exponential cooling, although not clearly distinguishable from the 
power-law on the grounds of the $SN1987A$ data, 
predicts at later times (\mb{t\gapprox 20\: s}) a substantially reduced 
$\anue$ flux (see again Figure 12). This behaviour can account for the
sensitivity to smaller masses: in fact, with a lower $\anue$ background, 
the presence of delayed massive neutrinos is much more evident.

Actually, there is no physical reason for assuming an exponential law, and 
theoretical models generally predict a slower decrease of neutrino luminosity 
at late times \footnote{We remind that cooling of a degenerate gas by 
black-body radiation yields a power-law decrease of temperature: \mb{T(t)=T^0 
\left( 1+{t\over \tau}\right) ^{-1/2}}.}. Furthermore, when compared with 
$SN1987A$ data, exponential cooling models yield fits which are worse than 
those of power-law models \cite{b&s}. In conclusion, it seems to us that 
exponential cooling models are not very reliable for predicting late times 
neutrino luminosities.

\sez{Comparison with Krauss's results.}

Krauss {\it et al.} \cite{krauss}, once developed their 
analysis method, also evaluated the minimal detectable mass. They found 
that a mass as small as 50 $\ev$ is detectable at 99\% C.L. 
with Super-Kamiokande, for a medium luminosity burst from a SN at $D=10\, kpc$. 

This result is more optimistic than we found by using Burrows's
theoretical models and/or $SN1987A$ phenomenological models (we recall, e.g.,
that for Burrows's most favourable model one has \mb{m_{3\sigma}= 120\:\ev} at
99.7\% C.L.).

This discrepancy can't be ascribed to the differences in the analysis 
procedures. We have verified that, with our approach, one obtains
for the middle model of \cite{krauss} (n. 17) a minimal detectable mass 
\mb{m_{3\sigma}\simeq 48\:\ev}.

Actually, in the supernova models of \cite{krauss},
after an initial accretion phase (\mb{t\sim 0.5\: s}), the
luminosities of all neutrino flavors are assumed to decay {\it exponentially
with a very fast rate} (\mb{\tau _L\sim 1\: s}). This results in a quite
short tail of the event distribution, and so the signal of delayed $\nut$ is
much more evident, as we already noted for the $SN1987A$ exponential model.

This feature is evident if we compare a typical model from 
\cite{krauss} with one of those we used in the foregoing.
In Figure 13, for instance, we plotted the isotropic (scaled to $\thf$) and 
directional event rates both in Krauss's model 17 and in Burrows's model 54,
which give fairly the same event number. As we can see, with just 
\mb{m=100\,\ev}, in Krauss's model the directional signal outruns the 
isotropic ``background'' after a few seconds, while in Burrows's model this 
occurs only at a very later times (\mb{t>30\: s}). 

It looks to us that the fast luminosity decay assumed in \cite{krauss} is not 
justified since: $i)$ phenomenological models with exponential cooling yield
longer decay times when adapted to fit $SN1987A$ data, see \cite{l&l,b&s};
$ii)$ also theoretical models predict longer decay times, see \cite{burr88} 
and the previous discussion.

\begin{figure}[p]
\centering\leavevmode
\epsfbox{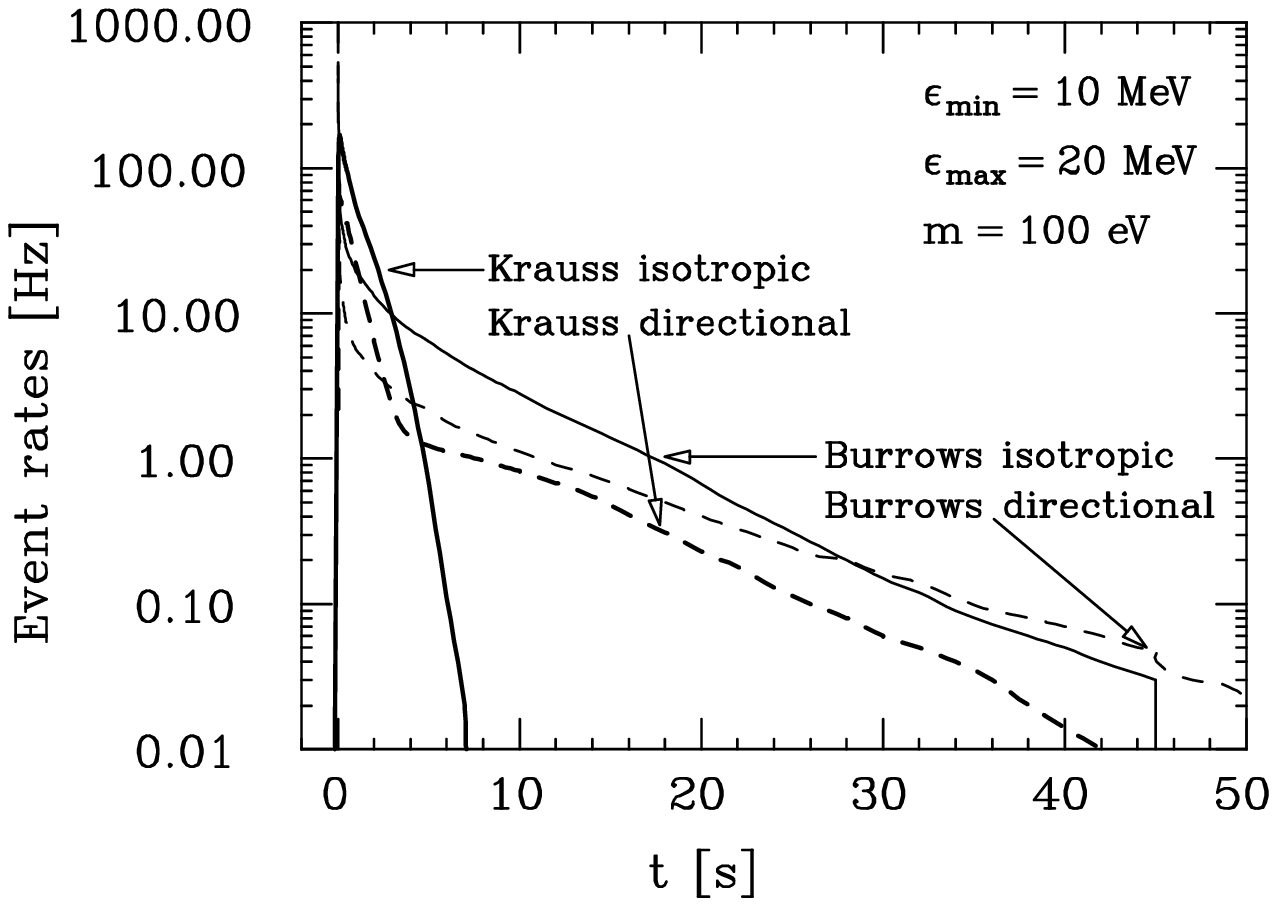}
\cap{Isotropic (scaled to $\thf$) and directional signals
in Krauss's model n. 17 and in Burrows's model n. 54 (same event number); 
a $m =100\,\ev$ was assumed.}
\end{figure}

\sez{Upper bounds.}

The situation may look somehow discouraging; nevertheless we should
remember that the present experimental limit to $\nut$ mass is tens $\mev$.
Therefore even in the most probable case of non-observation of mass effects, 
we will however significantly lower this limit. 

In order to obtain actual upper bounds, anyway, we have to introduce a more 
stringent statistical criterion than the one we used to define our 
``detectable'' mass.
With our definition of Section 2.4, in fact, we calculated the minimal value
of $m$ expected to produce a $3\sigma$ effect, checking against statistical
fluctuations of the massless case. We should now take into account
fluctuations of both massive and massless case.

Therefore, the new best indicator is the quantity
\equation \bar S (m) \equiv \max _{[t_1,t_2],[\emin,\emax]} 
{\abs{N_m - N_0} \over \sqrt{N_m + N_0}} \endequation
As an example, if \mb{\bar S(100\: \ev) =3} we can claim the upper 
bound $m\le 100\: \ev$ at a 3$\sigma$ level, i.e. with a 99.7\% 
Confidence Level.

In practice, $N_m$ is generally (much) larger than $N_0$, so that (6.1) can be 
replaced to a good approximation with:
\equation \bar S (m) \equiv \max _{[t_1,t_2],[\emin,\emax]} 
{\abs{N_m - N_0}\over \sqrt{N_m}} \endequation
This can be read very much like (2.15), i.e. assuming that neutrinos are 
massless, the observer will detect $N_0$ events (or a number close to it).
Eq. (6.2) tells then how many sigma's the result is out of the expectation
for massive neutrinos.

\begin{table}[h] 
\begin{center}  
\begin{tabular}{||l|c||c|c|c||}  
\hline  
\hline 
 & & & & \\[-.8em] 
Model & $m_{3\sigma}$ [\ev] & $[\emin,\emax]$ [$\mev$] & $[t_1,t_2]$ [$s$] & 
$\bar m_{3\sigma}$ [\ev]\\[.4em] 
\hline
\hline
Burrows RE & 140 & [15,100] & [15,100] & 200 \\
Burrows MF & 120 & [10,20] & [15,95] & 140 \\
Power-law & 110 & [10,20] & [10,85] & 140 \\
Exponential & 70 & [10,20] & [10,60] & 90 \\
\hline
Krauss 17 & 50 & [10,20] & [5,15] & 70 \\
\hline
\hline
\end{tabular}
\cap{The minimal excludable mass for a few significative models of supernova.}
\end{center}
\end{table}

The results for two significative theoretical models (Burrow's reference and
most favourable models - see Section 3) and two phenomenological models 
(exponential and power-law cooling - see Section 4) are listed in Table 6.
Using the same notation of Section 2.3, we define the minimal excludable mass 
as \mb{\bar m_{3\sigma}\equiv \min \{ m\, :\, \bar S(m)\ge 3\} }.
Clearly a larger mass is required to probe an upper bound rather than a
simply detectable effect, for one has \mb{N_m > N_0 \:\Rightarrow\: 
\bar S(m) < S(m)}. 
Anyway, one can see that even in the case of non-observation of 
mass effects, we will able to lower the present upper bound on $\nut$ mass by 
5 orders of magnitude!

One should be aware, anyway, that the bounds of Table 6 (and also the previous 
results on the minimal detectable mass) may not hold if the
$\nut$ mass is too large. In this case, in fact, one has a too
broad event distribution, see for istance \cite{wolf}.
A very crude estimate of the maximum mass $m_{\it max}$ explorable by 
delay-based analyses can be obtained by requiring that the 
rate of delayed $\nut$ events keeps greater than the background rate.
For SK one finds roughly \mb{m_{\it max}\sim {\cal O}(10\:\kev)}. Our results 
should be read therefore as lower extrema of an excluded mass region: in 
the case of non-observation of mass effects one could exclude $m$ in the 
window \mb{200\: \ev - 20\:\kev}.

\sez{Conclusions.}

Let us summarize the main points of this paper:
\begin{list}{\roman{enumi})}{\usecounter{enumi}}
\item We performed an extensive investigation of the sensitivity to 
non-vanishing $\nut$ mass in a large water \v{C}erenkov detector, developing
the analysis method introduced by Krauss {\it et al.} \cite{krauss}.
As the most important point is the dependence of the results on the supernova 
model, we analysed several theoretical models from numerical simulations of 
Burrows \cite{burr88} and phenomenological models based on $SN1987A$ data 
\cite{b&s,l&l}.
\item We determined optimal values of the analysis parameters (energy cuts and 
time window) so as to reach the highest sensitivity to a non-vanishing $\nut$ 
mass.
\item Our conclusion is that the minimal detectable mass is generally just 
above the cosmologically interesting range, $m\sim 100\:\ev$, see Table 2 
for details.
\item Differences with respect to \cite{krauss}, which claimed sensitivity to 
significantly smaller masses, are ascribed to the shorter decay time of 
neutrino luminosities assumed in \cite{krauss}, contrary to theoretical models 
as well to $SN1987A$ phenomenological models.
\item For the case that no positive signal is obtained, observation of a 
neutrino burst with SK will anyhow lower the present upper bound on $\nut$ mass
(\mb{m\lapprox 24\:\mev} \cite{pdg}) to few hundred $\ev$, see Table 6.
\end{list}

\appendix

\sez{Appendix}

\subsection{Super-Kamiokande characteristics.}

In Table 7 we summarize the main characteristics of the Super-Kamiokande
detector (the fiducial mass, the threshold energy and detection efficiency, and
the background rate spectrum) as we used in our analysis.

\begin{table}[h]
\begin{center}
\begin{tabular}{|cclr|}
\hline
\hline
 & & & \\[-.8em] 
$M_{det}^\dagger$ & = & $32\; kton$ & \\
$\epsilon _{th}$ & = & $5\; \mev$ & \\
$\eta(\ee)$ & $\simeq$ & $\vartheta (\ee - \epsilon _{th})\cdot
\left[ 0.93-e^{-(\ee\: /\: 9\, \mev)^{2.5}}\right]$ & \cite{burr88} \\
$b(\epsilon)$ & $\simeq$ &  $M_{det}\cdot {0.082\over \sqrt{2\pi}\, 0.87}\,
e^{-{1\over 2}\left( {\epsilon - 6.2 \over 0.87}\right) ^2}\: {\#\over \mev\, 
s}\;$ & elab. \cite{l&l} \\[.4em]
\hline
\hline
\multicolumn{4}{c}{$^\dagger$ We refer to the fiducial mass for supernova 
events.}
\end{tabular}
\end{center}
\cap{Super Kamiokande characteristics.}
\end{table}

\subsection{Cross sections.}

\begin{enumerate}
    \item \it capture on a proton: \rm
\equation {d\sigma _{\anue p}\over d\ee}={1\over 4}\sigma _0 \left( {\eps \over
m_e c^2}\right) ^2 \: (1 + 3\alpha ^2)(1+\delta _{wm}) \left( 1 -{Q \over \eps}
\right) \left[\left(1-{Q \over \eps} \right)^2 -\left({m_e c^2 \over\eps}\right)
^2 \right]^{1/2} \cdot \delta(\eps - \ee -Q) \endequation
where $\alpha \simeq -1.26$ is the axial-vector coupling constant, $Q\simeq
1.293\,\mev$ is the neutron-proton mass difference, \mb{\delta _{wm}\simeq
-3.3\cdot 10^{-3}(\eps -Q/2)/\mev} is the weak-magnetism correction,
and
$$\sigma _0 \equiv {4\over\pi}{1\over(c\hbar)^4} G_F^2 (m_e c^2)^2 \simeq 1.76
\cdot 10^{-44} \; cm^2$$
    \item \it scattering off an electron: \rm
\equation  {d\sigma _{\nu_x e} \over d\ee} = {1\over 2}{\sigma _0 \over m_e c^2}
\left[ A_x + B_x \left( 1-{\ee \over \eps}\right) ^2 -{m_e c^2 \ee \over\eps ^2}
\sqrt{A_x B_x} \right] \endequation
where the constants $A$ and $B$ have values:
$$ A_x = \left\{ \begin{array}{lllll}  \left( {1\over  2}+\sin  ^2  \theta  _W
\right) ^2 & \simeq & 0.536 &\;  ,  \hskip 10pt & \nu_e \\ \sin ^4 \theta _W &
\simeq & 0.0538 &\;  ,  \hskip 10pt & \bar \nu_e \\ \left( -{1\over 2}+\sin ^2
\theta  _W \right) ^2 & \simeq & 0.0719 &\;  ,  \hskip 10pt & \nu_i \\ \sin ^4
\theta _W & \simeq & 0.0538 &\;  ,  \hskip 10pt & \bar  \nu_i  \\  \end{array}
\right. $$
$$  B_x = \left\{ \begin{array}{lllll} \sin ^4 \theta _W & \simeq & 0.0538 & ,
\hskip 10pt & \nu_e \\ \left({1\over 2} +\sin ^2 \theta _W \right) ^2 & \simeq
& 0.536 & ,  \hskip 10pt & \bar \nu_e \\ \sin ^4 \theta _W & \simeq & 0.0538 &
,  \hskip  10pt  &  \nu_i \\ \left(-{1\over 2} +\sin ^2 \theta _W \right) ^2 &
\simeq & 0.0719 & , \hskip 10pt & \bar \nu_i \\ \end{array} \right. $$
    \item \it capture of a $\nue$ on an oxygen nucleus: \rm \\
The cross section of this process can be roughly approximated to \cite{nue_ox}
\equation {d\sigma _{\nue O} \over d\ee} \simeq 0.16\sigma _0 \left( {\eps -
\tilde \epsilon \over m_e c^2}\right) ^2 \theta (\eps - \tilde \epsilon)\,\cdot
\, \delta (\eps - \tilde \epsilon -\ee + m_e)\endequation
where $\tilde \epsilon \sim 13\,\mev$ is an effective reaction threshold.
    \item \it capture of a $\anue$ on an oxygen nucleus: \rm \\
We use the approximation given by \cite{krauss}
\equation {d\sigma _{\anue O} \over d\ee} \simeq 0.074\sigma _0 \left( 
{\eps - \tilde \epsilon \over m_e c^2}\right) ^2 \theta (\eps -\tilde \epsilon)
\,\cdot\, \delta (\eps - \tilde \epsilon -\ee + m_e)\endequation
where the reaction threshold is again $\tilde \epsilon \sim 13\,\mev$.
\end{enumerate}

\subsection{From emission to detection.}

Being \mb{d^2 N_{\nux} / d\eps\, dt'} the rate spectrum of $\nux$ emitted  
at time $t'$ with energy $\eps$, as defined in Section 2.3, the corresponding
differential flux at the detector at time $t$ is
\equation {d^2 \Phi_{\nux}(\eps ,t) \over d\eps\, dt}={1\over 4\pi D^2}\int dt'
\, {d^2 N_{\nux}(\eps ,t') \over d\eps \, dt'}\,\cdot\,\delta (t-t'-\Delta 
t(\eps, m))\endequation

For each detection process one obtains therefore the event rate $dN_b /dt$ as
\equation {dN_b (t)\over dt} = \int _{\emin} ^{\emax} d\ee \,\eta 
(\ee) \int _0 ^\infty d\eps {d^2 \Phi_{\nux}\over  d\eps \, dt}\, n_b\,
{d \sigma _{\nu b} (\eps ,\ee ) \over d\ee}\endequation
where \mb{d\sigma _{\nu b} / d\ee} is the differential cross section
of the reaction under examination, $n_b$ is the target number, and 
\mb{\eta (\ee)} is the electron/positron detection efficiency.

For the natural background one has instead the time-independent rate
\equation  B(\emin,\emax) =\int _{\emin} ^{\emax} d\ee\, b(\ee)\, \eta (\ee)
\endequation
where $b(\ee)$ is the empirically determinated background rate spectrum.


\end{document}